\documentclass{appolb}
\usepackage{epsfig}

\def\er#1#2{\relax\ifmmode{}^{+#1}_{-#2}\else$^{+#1}_{-#2}$\fi}
\newcommand{\be}{\begin{equation}}
\newcommand{\bea}{\begin{eqnarray}}
\newcommand{\ee}{\end{equation}}
\newcommand{\eea}{\end{eqnarray}}

\def\slash#1{\setbox0=\hbox{$#1$}
   \dimen0=\wd0 \setbox1=\hbox{/} \dimen1=\wd1
   \ifdim\dimen0>\dimen1 \rlap{\hbox to \dimen0{\hfil/\hfil}} #1
   \else  \rlap{\hbox to \dimen1{\hfil$#1$\hfil}} / \fi}

\preprint{} 
\begin{document}
\date{\today} 
\title{PION STRUCTURE AT HIGH AND LOW ENERGIES IN CHIRAL QUARK MODELS
\thanks{Lectures given at the XLII Krakow School on Theoretical
Physics (Zakopane), May 31st-June 9th 2002}}
\author{E. Ruiz Arriola~\footnote{e-mail:earriola@ugr.es}
\address{Departamento de F\'{\i}sica Moderna. Universidad de
Granada. \\ 
E-18071 Granada.  Spain.}} 
\maketitle



\begin{abstract}
Low and high energy properties of the pion are reviewed in the
framework of chiral quark models. Particular emphasis is put on the
simplest version of the SU(2) NJL model as prototype. The role of
gauge invariance in this kind of calculations is stressed. The results
are used as initial conditions for perturbative QCD evolution
equations. At leading order the quark model scale is $\mu_0 \sim 320
{\rm MeV} $ as determined from the pion distribution functions and the
pion distribution amplitudes.
\end{abstract}


\section{Introduction and Motivation} 

The structure and dynamics of hadrons is known to be a difficult
problem because the underlying QCD theory in terms of quark and gluon
degrees of freedom is actually very complicated~\cite{Yn83}. There
are, however, well known situations where some helpful simplifications
take place.

At high energies asymptotic freedom sets in, and perturbative QCD
becomes applicable. The renormalization group equations allow, on the
basis of the operator product expansion (OPE), to relate hadronic
matrix elements of quark and gluon operators at different
scales~\cite{Pe79}. There are many situations where this scale
dependence has been predicted and in some cases tested, particularly
in the study of high energy inclusive~\cite{AP77} and exclusive
processes~\cite{BL80}, but perturbative QCD says nothing about the
value of these matrix elements at a given scale.

At very low energies, spontaneous chiral symmetry breaking of the QCD
Lagrangian dominates, and the would-be Goldstone bosons in the massless
quark limit become the lightest states in the hadronic spectrum. They
are naturally identified with the pion, and it can be shown that at
low energies the interaction among pions becomes small. Actually, at
energies of the order of the pion mass, Chiral Perturbation Theory
(ChPT) can be applied, in terms of low energy constants which
correspond to certain QCD Green's functions, and many successful
results have been found~\cite{GL84,Pi95}. The values of the low energy
constants cannot be computed themselves from ChPT.
   
The previous considerations suggest the study of pion properties at
high energies, be exclusive or inclusive processes, trying to exploit
Chiral Symmetry. From a theoretical viewpoint, $\pi$ mesons are
particularly distinguished hadrons since most of their low energy
properties follow the patterns dictated by chiral symmetry. Actually,
we do not expect to understand the properties of any hadron better
than the pion, as Chiral Perturbation Theory suggests. Recent studies
have shown, for instance, that ChPT can be applied to compute chiral
corrections of hadronic matrix elements involved in deep inelastic
scattering~\cite{AS01} (in much the same way as chiral corrections are
computed for $\pi\pi$ scattering). However, nothing is said on the
value of those matrix elements in the chiral limit.

The calculation of pion matrix elements of quark and gluon operators
involved in high energy exclusive and inclusive processes requires
some non-perturbative method. Along these lectures it will be shown
how this can be done with the help of low energy chiral quark models,
which account for spontaneous chiral symmetry breaking, supplemented
with QCD evolution equations, which automatically implement QCD
logarithmic radiative corrections.

Chiral Quark Models share with QCD several features, mainly chiral
symmetry (realized in the Goldstone phase) and explicit quark degrees
of freedom. As we will see below they are useful in that they can make
quantitative predictions which, in many cases, agree rather well with
experiment. Unfortunately, there is not an unique model, but many and
almost as many versions as authors working on the field. So, for
practical reasons we will be using the Nambu--Jona-Lasinio
model~\cite{NJL61} for quark degrees of freedom~\cite{NJL}. It turns
out that, when dealing with low energy properties of the pion, model
details seem rather irrelevant, so most people have not paid
particular attention to their different versions.  This seems not to
be the case for properties which are extracted from high energy
inclusive and exclusive processes where the model details and/or their
different versions turn out to produce quite different results.

Given the fact that QCD is a theory with quarks and gluons, one
obvious question one should ask is what is the scale at which a
(chiral) quark model is naturally defined taking into account that its
only explicit degrees of freedom are quarks. So far, we only know of
one possible quantitative answer~\cite{JR80,Ja85}, namely the adequate
scale is defined by requiring that quarks carry all momentum of the
hadron. In a field theoretical chiral quark model this is a trivial
statement, because there are no explicit gluons. This assumption seems
consistent with QCD in perturbation theory since it turns out that the
gluon contribution to the energy-momentum tensor becomes smaller as
the running scale goes down until it eventually vanishes. The
remaining quark contribution can be separated into sea (flavor
singlet) and valence (flavor non-singlet) pieces. The precise value
suggested by perturbative calculations is in the range, $\mu = 300-350
{\rm MeV} $. Of course, the separation of the gluon, sea-quark and
valence-quark momentum fractions is a not a renormalization group
invariant operation. Thus, if such a matching procedure between QCD
and the chiral quark is meaningful, it depends on the particular
renormalization scheme.

\section{Low energy models from high energies} 

QCD is an asymptotically free theory, which means that the universal
coupling constant among quarks and gluons becomes small as the renormalization
scale goes to infinity, and hence perturbation theory becomes
applicable.  For definiteness, we use LO evolution for the running
strong coupling constant~\cite{AP77},
\begin{eqnarray} 
 \alpha(\mu)= \left( \frac{4 \pi}{\beta_0 } \right) \frac1{\log (\mu^2
 / \Lambda_{\rm QCD}^2 )}
\end{eqnarray} 
with $ \beta_0 = 11 C_A /3 - 2 N_F /3 $, $C_A = 3$ and $N_F$ being the
number of active flavors, which we take equal to three. We take for
concreteness $\Lambda_{\rm QCD}=226~{\rm MeV} $, which for $\mu = 2
{\rm GeV} $ yields $\alpha=0.32 $~\cite{PDG}. In perturbative QCD
calculations, $\alpha / 2 \pi $ is the expansion
parameter. Renormalization group invariance requires a very specific
dependence for any set of observables, $O$, on the scale
\begin{eqnarray}
O (\mu) = U(\mu, \mu_0)  O (\mu_0 )
\end{eqnarray}  
where $U(\mu,\mu_0)$ is a linear matrix operator, fulfilling cocycle
properties typical of evolution equations,
\begin{eqnarray} 
U(\mu_1 , \mu_2 ) U(\mu_2, \mu_3) &=& U(\mu_1 , \mu_3) \\
U(\mu,\mu)&=& 1
\label{eq:cocycle} 
\end{eqnarray} 
In the limit $\mu , \mu' \to \infty $ the operators are explicitly
calculable in perturbation theory. Obviously, conserved quantities
do not depend on the evolution scale. 

A particularly interesting operator in QCD is the energy-momentum
tensor $\theta^{\alpha \beta}$, which due to Poincare invariance is a
conserved quantity and hence renormalization invariant. Its matrix
element between pion states fulfills, 
\begin{eqnarray}
\langle \pi | \theta^{\alpha \beta} | \pi \rangle = 2 p^\alpha p^\beta  
\end{eqnarray} 
For the QCD Lagrangian the energy momentum tensor can be separated
into three contributions~\cite{AP77},
\begin{eqnarray}
\theta^{\alpha \beta}= \theta_G^{\alpha \beta} + \theta_{q,{\rm
S}}^{\alpha \beta} + \theta_{q,{\rm NS}}^{\alpha \beta}
\end{eqnarray} 
where $\theta_G^{\alpha \beta}$,$\theta_{q,{\rm S}}$ and
$\theta_{q,{\rm NS}}^{\alpha \beta}$ are the gluon,quark--singlet and
quark-- non-singlet (or valence) contributions respectively fulfilling 
\begin{eqnarray}
\langle \pi | \theta^{\alpha \beta}_G | \pi \rangle |_\mu &=& 2
p^\alpha p^\beta G_1 (\mu) \nonumber \\ \langle \pi | \theta^{\alpha
\beta}_{q,{\rm S}} | \pi \rangle |_\mu &=& 2 p^\alpha p^\beta S_1
(\mu) \\ \langle \pi | \theta^{\alpha \beta}_{q,{\rm NS}} | \pi \rangle
|_\mu &=& 2 p^\alpha p^\beta G_1 (\mu) \nonumber
\end{eqnarray} 
where $G_1(\mu)$, $ S_1(\mu) $ and $V_1 (\mu)$ are the gluon, sea
quark and valence quark momentum fractions of the pion. In deep
inelastic scattering (DIS) it can be shown~\cite{AP77} that if
$q_i(x,\mu)$, $\bar q_i(x,\mu)$ and $G(x,\mu)$ represent the
probability density of finding a quark, antiquark and gluon
respectively with momentum fraction $x$ at the scale $\mu$ ($\mu^2 =
Q^2$ in DIS) then one has
\begin{eqnarray}
G_1 (\mu) &=& \int_0^1 dx x G(x,\mu)  \\
S_1 (\mu) &=& \int_0^1 dx x S(x,\mu)  \\
V_1 (\mu) &=& \int_0^1 dx x V(x,\mu)  
\end{eqnarray} 
where, for $\pi^+$, $ V=u_\pi - \bar u_\pi + \bar d_\pi - d_\pi $ and
$ S=u_\pi - \bar u_\pi - \bar d_\pi + d_\pi $. These momentum
fractions depend on the scale, $\mu$, and fulfill the momentum sum
rule
\begin{eqnarray}
V_1(\mu) + S_1(\mu) + G_1 (\mu) = 1  
\end{eqnarray} 
as a consequence of the energy-momentum tensor conservation. In
perturbation theory it is verified that due to radiative corrections
$G_1(\mu)$ and $S_1(\mu)$ decrease as the scale $\mu$ goes down. On
the contrary, the non-singlet contribution to the energy momentum
tensor evolves as
\begin{eqnarray}
\frac{ V_1 (\mu) } { V_1 (\mu_0) } = \left( \frac{\alpha(\mu)}
{\alpha(\mu_0) } \right)^{\gamma_1^{\rm NS} / 2 \beta_0 } \quad ,
\qquad
\end{eqnarray} 
where $ \gamma_1^{\rm NS} / 2 \beta_0 = 32/81 $ for $N_F=N_c=3$.  The
value of $V_1 (\mu)$ has been extracted from the analysis of high
energy experiments. In Ref.~\cite{SMRS92} it was found that at $Q^2 =
4 {\rm GeV}^2 $ valence quarks carry $47 \%$ of the total momentum
fraction in the pion, e.g., for $\pi^+$,
\begin{eqnarray} 
V_1 = \langle x \left( u_\pi - \bar u_\pi + \bar d_\pi - d_\pi \right)
\rangle = 0.47 \pm 0.02 \qquad {\rm at} \qquad \mu^2 = 4 {\rm GeV}^2
\; .
\end{eqnarray} 
Downwards LO evolution yields that for a given reference scale, $\mu_0$,
\begin{eqnarray}
V_1 (\mu_0) = 1 \qquad G_1 (\mu_0) + S_1 (\mu_0) = 0
\end{eqnarray} 
The scale $\mu_0$ so defined is called the quark model point for
obvious reasons. At LO the scale turns out to be
\begin{eqnarray}
\mu_0 = 313_{-10}^{+20} {\rm MeV}  
\label{eq:mu0_dis} 
\end{eqnarray} 
This is admittedly a rather low scale, but one can still hope that the
typical expansion parameter $\alpha (\mu_0) / 2 \pi \sim 0.34 \pm 0.04
$ makes perturbation theory meaningful. Since such an approach was
first suggested~\cite{JR80} that is all one can do for the
moment~\footnote{Actually, in the case of the nucleon these low scales
produce negative gluon densities, if one takes $ V_1 (\mu =2 {\rm
GeV}) =0.40 $ and hence violate positivity of parton distribution
functions as well as unitarity of structure
functions~\cite{Ru98a,Co01}.}. There are more sources of uncertainties
to Eq.~(\ref{eq:mu0_dis}). For instance, if the point for the quark
model is defined by $G_1 (\mu_G) =0$, then at LO the scale is $\mu_G
\sim 350 {\rm MeV}$. The determination of $\mu_0$ given by
Eq.~(\ref{eq:mu0_dis}) is model independent. In these lectures we will
show how this determination not only leads to a successful description
of non-singlet pionic parton distribution functions in certain
versions of the chiral quark model, but also that the number is in
quantitative agreement with other determinations.

\section{\bf Chiral symmetry and chiral quark models} 

For the SU(2) up and down quarks, chiral symmetry is the invariance of
the QCD Lagrangian under the transformations~\cite{Pi95},
\begin{eqnarray} 
q(x) && \to e^{ {\rm i} \gamma } q(x) \\ q(x) && \to e^{ {\rm i} \vec
\alpha \cdot \vec \tau} q(x) \\ q(x) && \to e^{{\rm i} \vec \beta
\cdot \vec \tau \gamma_5 } q(x)
\end{eqnarray} 
where $q(x)$ represent Dirac spinor fields with $N_F=2$ flavors and
$N_c$ colors where $\vec \tau $ are the isospin Pauli matrices. As a
consequence of this symmetry there exist baryon, vector and axial
Noether currents,
\begin{eqnarray}
J_B^{\mu} (x) &=& \bar q(x) \gamma^\mu q(x) \qquad \qquad \qquad {\rm
Baryon \quad  current }
\label{eq:BC}
\\ J_V^{\mu,a} (x) &=& \bar q(x) \gamma^\mu
{\tau_a \over 2} q(x) \qquad \qquad \quad {\rm
Vector \quad  current }
\label{eq:VC}
 \\ J_A^{\mu,a} (x) &=& \bar q(x) \gamma^\mu
\gamma_5 {\tau_a \over 2} q(x) \qquad \qquad {\rm
Axial \qquad  current }
\label{eq:AC}
\end{eqnarray} 
respectively.  Conservation of the vector current (CVC) and partial
conservation of the axial current (PCAC) imply that
\begin{eqnarray}
\partial_\mu J_B^{\mu} (x) &=& 0 
\label{eq:CBC} \\
\partial_\mu J_V^{\mu,a} (x) &=& 0 \label{eq:CVC} \\ \partial_\mu
J_A^{\mu,a} (x) &=&  m \bar q(x)   i \gamma_5 \tau_a  q(x)
\label{eq:PCAC} 
\end{eqnarray} 
with $m$ denoting the average current quark mass (we neglect isospin
breaking effects). A chiral quark model is any chirally invariant
dynamical field theoretical model containing only explicit quark
degrees of freedom and fulfilling the previous conservation
laws~(\ref{eq:CBC}), (\ref{eq:CVC}) and
(\ref{eq:PCAC}). Unfortunately, there is no such a thing as {\it the}
chiral quark model. This is probably the reason why there exists a
proliferation of models with many variants. Regardless of their
differences all these models are characterized by the three ways of
breaking chiral symmetry in QCD: explicit, spontaneous and anomalous.
These are non trivial requirements, which imply constraints on the
regularization methods in local models or equivalently on the high
energy behavior of non-local models.  In fact non-local models have
been preferred because they provide a more natural explanation of the
anomalous $\pi^0 \to \gamma \gamma $ decay rate~\cite{BR93}, but the
calculations are cumbersome. In addition, they are formulated in
Euclidean space and their extrapolation to compute matrix elements
involved in high energy processes is subtle as it is on the
lattice. For instance, the calculation of structure functions requires
either a continuation of the non-local model to Minkowki space or the
determination of some moments in the Euclidean region and subsequent
distribution reconstruction. While the first possibility turns out to
be extremely difficult to explore, the second alternative cannot be
used to pin down the $x \to 1$ which depends on the asymptotic
behavior of the moments. On the contrary, local models, although less
realistic, require specifying a suitable regularization which may be
directly formulated in Minkowski space.

\subsection{NJL Lagrangian} 

For the purpose of our discussion we think it is useful to review the
SU(2) NJL model~\cite{NJL61,NJL}. This model has extensively been used
in the past and there exist many reviews on the subject. So, although
we aim at a self-contained discussion of pion properties, our
presentation will be necessarily sketchy in order to provide the main
ideas. Nevertheless, we will stress those points where differences
with other authors become important. The NJL Lagrangian in Minkowski
space is given by \cite{NJL61,NJL}
\begin{eqnarray}
{\cal L}_{\rm NJL} &=&
\bar{q} (i\slash\partial - m )q  +
{G \over 2} \left( (\bar{q}  q)^2
                        +(\bar{q}\vec \tau i \gamma_5 q)^2 \right)  
\label{eq:NJL} 
\end{eqnarray}
where $q(x)$ is a Dirac spinor field with $N_c=3$ colors and $N_F=2$
flavors, $G$ is the coupling constant with dimension $[G]=M^{-2}$ and
$m$ the average mass of the current up and down quarks $(\sim 7 {\rm
MeV} $ at the scale $\mu \sim 1 {\rm GeV} )$.  With the exception of
the term in the current quark mass $m$ , this Lagrangian is invariant under
the $U_B (1) \otimes SU(2)_R \otimes SU(2)_L $ chiral group, with a
$U(1)_B \otimes SU(2)_V $ subgroup. Thus, the currents (\ref{eq:BC}),
(\ref{eq:VC}) and (\ref{eq:AC}), and their conservation laws
(\ref{eq:CBC}), (\ref{eq:CVC}) and (\ref{eq:PCAC}) are satisfied, as
in QCD. A consequence of Poincar\'e invariance is the conservation of
the energy momentum tensor,
\begin{eqnarray}
\theta^{\mu \nu } = \frac{\rm i}2 \bar q \left( \gamma^\mu
\partial^\nu + \gamma^\nu \partial^\nu \right) q - g^{\mu\nu} {\cal L}
\end{eqnarray} 
which only contains quark degrees of freedom. Thus, we expect in this
model that for any hadron, and in particular for the pion  
\begin{eqnarray}
\langle \pi| \theta^{\mu \nu} | \pi \rangle = 2 p^\mu p^\nu \, . 
\end{eqnarray}  
Since the coupling constant, $G$, has a negative energy dimension, the
Lagrangian (\ref{eq:NJL}) above is not renormalizable by power
counting and it is usually interpreted as a theory with a finite
cut-off, which we generically denote by $\Lambda$. This means that
results depend crucially on the value of $\Lambda$ and some fine
tuning will be invoked. The need of a cut-off is not a serious problem
from a physical point of view because we know that at high energies,
much above the cut-off, the effective interaction should be replaced
by the underlying QCD interactions in terms of explicit quarks and
gluons. Roughly speaking, this feature is present in more
sophisticated chiral quark models with non-local interactions.

The real problem derived from the high-energy suppression has to do
with the fact that we want to preserve the symmetries of the
Lagrangian, namely gauge and chiral symmetry, without removing the
cut-off by taking the limit $ \Lambda \to \infty $, i.e. keeping the
high energy suppression. This implies that not every regularization,
although it may accommodate some particular prejudices, may be
considered as suitable. This point will be illustrated below. A full
discussion on regularization methods in the NJL model can be found in
Ref.~\cite{Do92}.

In these lectures we will work, as usual, in the one-quark loop
approximation. This is equivalent to work to leading order in the
large $N_c $ expansion, taking $ G N_c =$constant. In this way we
comply not only with chiral symmetry constraints, but also with large
$N_c$ requirements. The study of higher orders in a $1/N_c$ expansion
can be traced from Refs.~\cite{DS95,NB96,OB01}.

\subsection{Pauli-Villars regularization} 

In a cut-off theory most of the discussion is related to the
regularization, and hence gets a bit technical. We will use throughout
these lectures the time honored Pauli-Villars (PV) regularization
method~\cite{PV49} which has already been used for several
applications in the NJL model, like ChPT~\cite{Ru91}, finite
temperature~\cite{CR91}, solitons~\cite{SR92}, structure
functions~\cite{DR95,WRG99,DR02}, correlation functions~\cite{DR96}
and distribution amplitudes~\cite{RB02}. To our knowledge this is the
most reliable regularization method of the NJL model so far fulfilling
many desirable features, among others preserving gauge invariance and
working directly in Minkowski space (see the recent discussion in
Ref.~\cite{DR02}). The only difference with the standard PV method of
QED is that in the NJL model this type of regularization is applied at
the constituent quark mass level (see below). It is surprising that
given the advantages of the method, it has been used so few times as
compared with other methods such as Euclidean $O(4)$ cut-off or
Schwinger's Proper-Time. A potential disadvantage of the method is
that it provides a negative spectral strength due to the PV
subtractions, and thus positivity of some physical quantities, like
form factors or distribution functions might not be
fulfilled. Although this is not exclude, in all the calculations we
present we explicitly see that is not the case.

In the NJL model this method has been used mainly within the context
of the bosonization or auxiliary field method (see
e.g. Ref.~\cite{SR92}), but at the one loop level and for the
processes we will be considering it corresponds to making the
replacement under the momentum integral
\begin{eqnarray}
\frac1{\slash{k}_1-M}  \cdots  \frac1{\slash{k}_N-M} \to
\sum_i c_i \left\{ \frac{\slash{k}_1+M}{k_1^2 -M^2 - \Lambda_i^2}
\cdots \frac{\slash{k}_N+M}{k_N^2 -M^2 - \Lambda_i^2}
\right\}
\end{eqnarray} 
$c_i$ and $\Lambda_i$ are the same suitable coefficients for all
one loop graphs, fulfilling $c_0 = 1 $ and $\Lambda_0=0$ and chosen in
such a way as to make one loop integrals finite. In the graphs with an
odd number of Dirac $\gamma_5$'s the regulator has to be removed at
the end of the calculation. 

In the NJL there appear quadratic and logarithmic ultraviolet
divergences. To illustrate the method let us consider the
integral 
\begin{eqnarray}
I = -{\rm i}  \int \frac{d^4 k}{(2\pi)^4}
\frac1{k^2 -M^2}
\label{eq:I}
\end{eqnarray} 
which is quadratically ultraviolet divergent, as can be seen by
integrating first over $k_0$ component (using Feynman's $M^2 \to M^2 -
{\rm i} 0^+ $ prescription) and then over the $\vec k$
component. Using the PV subtractions, it becomes 
\begin{eqnarray}
I_{\rm PV} = -{\rm i}  \sum_i c_i \int \frac{d^4 k}{(2\pi)^4}
\frac1{k^2 -M^2-\Lambda_i^2 }
\label{eq:I_PV} 
\end{eqnarray} 
Evaluating the integral by any method, one sees that at least the
following two conditions are required to render the one loop Feynman
integrals finite
\begin{eqnarray}
\sum_i c_i = 0 \qquad \sum_i c_i \Lambda_i^2 = 0
\label{eq:PV}
\end{eqnarray} 
Thus, at least two subtractions are needed. Solving the system of
equations in terms of $\Lambda_1$ and $ \Lambda_2 $ provides two
unknown parameters. To reduce them to only one the coincidence limit $
\Lambda_1 \to \Lambda_2 = \Lambda$ is taken yielding the rule
\begin{eqnarray}
\sum_i c_i f( \Lambda_i^2 ) = f(0) - f(\Lambda^2 ) + \Lambda^2 f'
(\Lambda^2 )
\label{eq:PV2} 
\end{eqnarray} 
This calculation already illustrates a very general feature of the PV
method, calculations may directly be done in Minkowski space, and
there is no need to go to Euclidean space. This is a computational
advantage in the study of high energy processes as we will see below.
To simplify the notation we will implicitly assume the use of such a
regulator in what follows.

\subsection{Chiral symmetry breaking} 

\begin{figure}
\begin{center}
\epsfig{figure=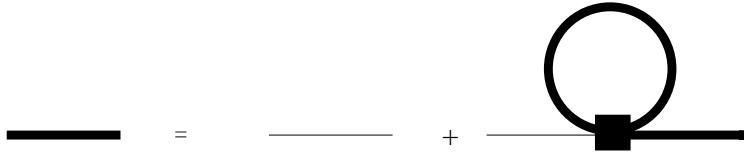,height=2cm,width=10cm} 
\end{center} 
\caption{Schwinger-Dyson equation for the quark propagator. The thick
line represents the full propagator, whereas the thin line stands for
the free quark propagator. The full blob is the irreducible two
particle amplitude.}
\label{fig:dyson}
\end{figure}

To illustrate the method let us consider the Dyson-Schwinger equation
for the quark propagator, $S(p)$, (see Fig.~\ref{fig:dyson}). 
\begin{eqnarray}
S(p) = S_0 (p) + S (p) (-{\rm i} ) \int \frac{d^4 k}{(2\pi)^4} {\rm
Tr} \left[ S(k) G \right] S_0 (p)
\end{eqnarray} 

where $S(p)$ and $ S_0(p) $ are the full and free quark propagators
respectively. $G$ is the four-point kernel, which in the lowest order
approximation becomes the coupling constant and can be taken out of
the integral. Thus the full propagator acquires a constant
self-energy, $\Sigma(p)=M$. Computing the Dirac, flavor and color
traces one gets the so-called gap equation in the convenient form
\begin{eqnarray} 
M \left( 1 - 4 G N_F N_c I \right) = m 
\label{eq:gap} 
\end{eqnarray} 
where Eq.~(\ref{eq:I_PV}) is understood for the integral $I$. For
$m=0$ there are two solutions (the Wigner alternative)
\begin{itemize} 
\item Wigner phase. Corresponding to $M=0$. 
\item Goldstone phase. $M \neq 0 $ and hence 
\begin{eqnarray} 
1 = 4 G N_F N_c I
\label{eq:gap0} 
\end{eqnarray} 
which is called the gap equation in the limit $m \to 0$ because it
provides a gap of width $2 M $ in the quark spectrum. 
\end{itemize} 
For $m \neq 0 $ the gap equation is Eq.~(\ref{eq:gap}) and one has to
choose the solution which goes to Eq.~(\ref{eq:gap0}) for $m \to 0$. 

\subsection{Quark condensate} 

In the chirally broken phase, one has a quark condensate ( $ \langle
\bar q q \rangle \equiv \langle \bar u u + \bar d d \rangle $)
\begin{eqnarray}
\langle \bar q q \rangle =
(-{\rm i} ) {\rm Tr} \int \frac{d^4 k}{(2\pi)^4} \frac1{\slash{k}-M} =
-8 N_c M I = -\frac{M-m}G \neq 0
\label{eq:qqbar}
\end{eqnarray}  
The standard accepted value in from QCD sum rules~\cite{CK01} is
$\langle \bar q q \rangle_\mu = \langle \bar u u + \bar d d
\rangle_\mu = -2 (240 \pm 10 \, {\rm MeV})^3 $ at the scale $\mu = 1 {\rm
GeV}$. At leading order in the QCD evolution one has
\begin{eqnarray}
\frac{ \langle \bar q q \rangle_\mu } {\langle \bar q q
\rangle_{\mu_0} } = \left( \frac{\alpha(\mu)} {\alpha(\mu_0) }
\right)^{\gamma_{\bar q q } / 2 \beta_0 } \quad , \qquad
\gamma_{\bar q q } = 8 \, .
\end{eqnarray} 
If we use the scale of Eq.~(\ref{eq:mu0_dis}) we get an enhancement
factor $ \sim 2 $ and hence, 
\begin{eqnarray}
\langle \bar q q \rangle_{\mu_0} = -(380 \pm 20 {\rm MeV})^3 . 
\end{eqnarray} 
According to our point of view it is this value the one that should be
compared to the model calculation, and not to the one at $\mu = 1 {\rm
GeV}$. Of course, one may argue that LO evolution is not sufficient to
go to such low scales. All we can do to check the consistency of the
approach is to provide, as we do below, other determinations of the
scale $\mu_0$. 

\begin{figure}[]
\begin{center}
\epsfig{figure=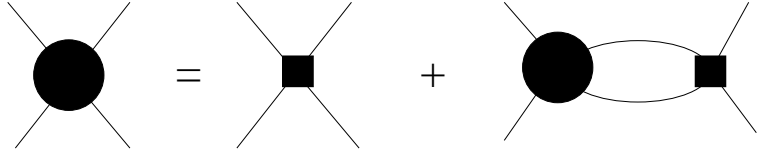,height=2cm,width=10cm}
\end{center} 
\caption{Bethe-Salpeter equation for quark-antiquark scattering. The
square blob represents the two particle irreducible quark-antiquark
potential, and the round blob is the full T-matrix. The lines
correspond to constituent quarks ( thick lines in
Fig.~(\ref{fig:dyson})).}
\label{fig:bethe}
\end{figure}

\subsection{The pion as quark-antiquark bound state} 

In the spontaneously broken phase, the constituent quark mass, $M \neq
0$. As a consequence of Goldstone's theorem there must exist massless
pseudoscalar particles, which one identifies with the pions. This can
actually be obtained by solving the Bethe-Salpeter equation in the
pseudoscalar-isovector channel for quark-antiquark scattering (see
Fig.~\ref{fig:bethe}) and checking that there is a bound state pole
which becomes massless as the current quark mass goes to zero, $m \to
0$. In the lowest order approximation, where the kernel is
approximated by a constant, the Bethe-Salpeter amplitude for $\pi^+$
becomes
\begin{eqnarray}
T_{\alpha \beta; \gamma \delta} = \left( \gamma_5 \tau^+
\right)_{\alpha \gamma} \left( \gamma_5 \tau^+ \right)_{\beta \delta}
t (P)
\end{eqnarray} 
where $ t (P) $ is a number depending only on the total momentum,
given by 
\begin{eqnarray}
t(P) = G  + G (-{\rm i} ) \int \frac{d^4 k}{(2\pi)^4}
{\rm Tr} \left[ \gamma_5 \frac1{\slash{k}-M} \gamma_5
\frac1{\slash{P}-\slash{k}-M} \right] t(P) 
\end{eqnarray} 
In writing the polarization operator, a particular choice of momentum
routing has been made, but in a gauge invariant regularization such as
PV, one may safely shift the integration variable by any amount of the
external momentum $P$. Euclidean $O(4)$ cut-offs do not preserve this
property. Using the PV regulators one gets
\begin{eqnarray}
t(p)^{-1} = G^{-1} - 8 N_c I +  4 N_c p^2 F(p^2)
\end{eqnarray} 
where $I$ is defined by Eq.~(\ref{eq:I}) and the one-loop function,
$F(p^2) $, is given by
\begin{eqnarray}
F(p^2) = (-{\rm i} ) \int \frac{d^4 k}{(2\pi)^4} \frac1{k^2 -M^2}
\frac1{(k-p)^2 -M^2}
\label{eq:F(p2)}
\end{eqnarray} 
Using the Feynman trick for the two propagators, this function can be
rewritten as
\begin{eqnarray}
F(p^2) &=& \int_0^1 dx F(p^2 , x ) 
\end{eqnarray} 
where 
\begin{eqnarray}
F(p^2, x) &=& (-{\rm i} ) \int \frac{d^4 k}{(2\pi)^4} \frac1{\left[k^2
-M^2 + x(1-x) p^2+ {\rm i} \epsilon \right]^2 } \nonumber \\ &=&
-\frac1{(4\pi)^2}\sum_i c_i \log\left[ M^2 + \Lambda_i^2 -x(1-x)p^2
\right]
\label{eq:Fpx}
\end{eqnarray} 
In the chiral limit, $m =0 $, the gap equation (\ref{eq:gap0}) can be
used, and the first terms in the r.h.s. cancel, producing a massless
quark-antiquark bound state. However, for $p^2 > 4 M^2 $, the function
$F(p^2) $, and hence the t-matrix, develops an imaginary part
indicating a lack of confinement. For $m_\pi^2 \ll 4 M^2 $ we get a
deeply bound state, and thus hope confinement not to be essential.
For $M = 300 {\rm MeV} $ one has $ m_\pi^2 / (4M^2) \sim 0.06 $.

\begin{figure}[tbp]
\begin{center}
\epsfig{figure=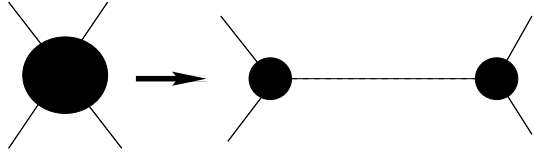,height=2cm,width=7cm}
\end{center} 
\caption{Pion pole contribution to the Bethe-Salpeter quark-antiquark
scattering amplitude.  The small solid blob represents the Bethe-Salpeter
pion quark-antiquark amplitude. Solid lines are constituent quarks.
.}
\label{fig:pion}
\end{figure}
For a finite quark mass we have a pole at $p^2 = m_\pi^2 $,
\begin{eqnarray}
\frac{m}{ G M} = 4 N_c m_\pi^2 F(m_\pi^2)
\label{eq:pion_pole}
\end{eqnarray} 
The pion coupling of a composite pion state to a quark-antiquark pair
is given by the residue of the $t$-matrix at the pion pole (see
Fig.~\ref{fig:pion} ),
\begin{eqnarray}
g_{\pi qq }^{-2} = \frac{d t^{-1} (p^2) }{d p^2 } \Big|_{p^2 =
m_\pi^2} = 4 N_c \frac{d \left[ p^2 F (p^2) \right] }{d p^2 }
\Big|_{p^2 = m_\pi^2}
\label{eq:gpqq} 
\end{eqnarray} 
So that the composite canonically quantized pion propagator is
\begin{eqnarray}
D(p^2) = \frac{t(p^2)}{g_{\pi qq}^2}  \to \frac1{p^2-m_\pi^2}   \quad
, \qquad 
( p^2 \to m_\pi^2 ) 
\label{eq:pion_prop} 
\end{eqnarray} 

\subsection{Pion weak decay}

\begin{figure}[tbp]
\begin{center}
\epsfig{figure=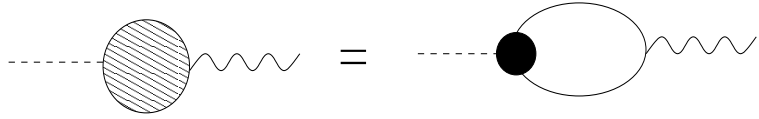,height=1.5cm,width=10cm}
\end{center} 
\caption{Pion (dashed line) weak decay into an axial current (wavy
line). The solid blob represents the Bethe-Salpeter
pion quark-antiquark amplitude. Solid lines are constituent quarks.
.}
\label{fig:fpi}
\end{figure}

Using the bound state solution of the Bethe-Salpeter equation for the
pion, one can compute the pion weak decay defined as 
\begin{eqnarray}
\langle 0 | J_A^{\mu,b} (0) | \pi^{a}(p) \rangle = {\rm i} f_\pi p^\mu
\delta^{ab}  
\end{eqnarray} 
yielding (see Fig.~\ref{fig:fpi}),
\begin{eqnarray}
f_\pi  = 4 N_c M g_{\pi qq} F(m_\pi^2)  
\label{eq:fpi} 
\end{eqnarray} 
and thus 
\begin{eqnarray}
g_{\pi qq } f_\pi = M \frac{ F(m_\pi^2) }{\left[ m_\pi^2 F (m_\pi^2)
\right]' }
\end{eqnarray} 
which, in the chiral limit yields the Goldberger-Treimans relation at
the quark level, $g_{\pi qq} f_\pi = M $. Combining
Eqs.~(\ref{eq:qqbar}), (\ref{eq:pion_pole}) and 
(\ref{eq:fpi})   one gets 
\begin{eqnarray}
- m \langle \bar q q \rangle = m_\pi^2 f_\pi^2 \left(
\frac{M-m}{g_{\pi qq} f_\pi } \right)
\label{eq:gor} 
\end{eqnarray} 
which at lowest order in the chiral expansion becomes the well known
Gell-Mann--Oakes-Renner relation.

Eqs.~(\ref{eq:fpi}) and (\ref{eq:gpqq}) are usually employed to fix
the parameters in the broken phase, taking for $f_\pi = 92 {\rm MeV} $
and $m_\pi = 139.6 {\rm MeV}$. For a given constituent quark mass,
$M$, one can compute the cut-off $\Lambda$. The gap equation allows to
determine the coupling constant $G$. For instance, using the PV
regularization given by Eq.~(\ref{eq:PV2}), for $M=280 {\rm MeV} $ one
obtains $\Lambda=871 {\rm MeV} $ and $\langle \bar q q \rangle = -(290
{\rm MeV} )^3 $. From this number one may deduce that, since it
resembles the numerical value of the condensate at the scale $\mu = 1
{\rm GeV}$, that the model scale is around $1 {\rm GeV}$. Of course,
the numerical values may change in a different version of the PV
scheme (like e.g. with $\Lambda_1 \neq \Lambda_2 $, or adding more
terms in the sum).  Fortunately, as we will see below, the
results for some high energy properties are insensitive to the choice
of parameters, up to unimportant chiral corrections.

\subsection{Pion wave function} 

In principle, the Bethe-Salpeter pion-quark-antiquark wave function is
given by (we take $\pi^+$)
\begin{eqnarray}
\chi_P (k) = \frac{\rm i}{\slash{P}/2+\slash{k}-M} \left(-g_{\pi qq}
\gamma_5 \tau^+ \right) \frac{\rm i}{\slash{P}/2-\slash{k}-M}
\label{eq:pion_wf}
\end{eqnarray} 
but PV regularization has to be understood because, at least formally, 
\begin{eqnarray}
f_\pi P_\mu = (-{\rm i}) \int \frac{d^4 k}{(2\pi)^4} {\rm Tr} \left(
\chi_P (k) \frac{\tau^+}2 \gamma^\mu \gamma_5 \right)
\end{eqnarray} 
and the integral is logarithmically divergent if
Eq.~(\ref{eq:pion_wf}) is taken literally. This problem is typical in
chiral quark models and has been the cause of much confusion.
Regularization is only easily applied at the level of closed quark
lines~\footnote{This is actually a good reason to use bosonization
schemes; there the concept of an open quark line does not appear
.}. We will see more of this later, but it has always been a weak
point in quark-loop calculations. One obvious choice to regularize
open lines is to enforce consistency with the closed lines, although
at first sight this procedure looks a bit arbitrary. Another
possibility to open a regularized quark line, but starting from a
closed quark line, is to consider a physical process involving photons
and pions in the high energy limit and extract the leading power
behavior. Prominent examples which will be analyzed below are deep
inelastic scattering from which the pion structure functions can be
deduced in the Bjorken limit or the pion transition $\gamma^* \to
\pi^0 \gamma $-form factor in the limit of high photon virtualities
and fixed photon assymetry from which the pion distribution amplitude
may be derived.

\subsection{Pion Electromagnetic Form Factor} 

\begin{figure}[tbp]
\begin{center}
\epsfig{figure=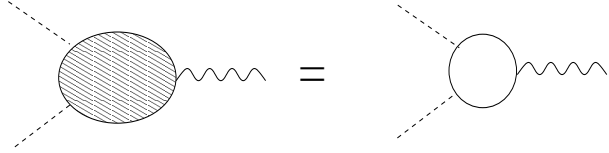,height=2cm,width=8cm}
\end{center} 
\caption{Electromagnetic pion form factor in the NJL model.  The
closed loops are Pauli-Villars regularized. Wavy lines are photons,
solid lines are constituent quarks and dashed lines composite pions.}
\label{fig:form_fac}
\end{figure}

The pion electromagnetic is probably the simplest case where one can
illustrate some of the points we want to make.  For a charged pion
$\pi^+ = u \bar d $ the electromagnetic form factor is defined as
\begin{eqnarray}
\langle \pi^+ (p') | J_\mu^{\rm em} (0) | \pi^+ (p) \rangle = e
\Gamma_\mu^{\rm em } (p' , p) = (p^\mu + p'^\mu) e F_\pi^{\rm em}
(q^2) 
\end{eqnarray} 
and can be computed using Fig.~\ref{fig:form_fac}.  For on-shell pions
and in the chiral limit the result is rather simple
\begin{eqnarray} 
F_\pi^{\rm em} (q^2) = \frac{4 N_c M^2 F(q^2)}{f_\pi^2}
\end{eqnarray} 
where $F(p^2)$ is given by Eq.~(\ref{eq:F(p2)}). The pion form factor
is obviously properly normalized, $ F_\pi^{\rm em} (0)= 1 $, as it
should be due to the gauge invariance of the Pauli-Villars
regularization. The mean-square pion radius reads
\begin{eqnarray}
\langle r^2 \rangle_\pi^{\rm em} = -6 {d F_\pi^{\rm em} (q^2) \over
dq^2 } \Big|_{q^2 =0} = {N_c \over 4 \pi^2 f_\pi^2} \times \sum_i c_i
\frac{M^2}{M^2 + \Lambda_i^2}
\end{eqnarray} 
a result explicitly depending on the regularization. Numerically one
has a reduction of about $25\%$ for $M=300 {\rm MeV} $ due to the
finite cut-off effect, $ ( \langle r^2 \rangle_\pi^{\rm em}) ^{1/2} =
(0.58 {\rm fm}) \times 0.75 = 0. 50 {\rm fm}$, to be compared with the
experimental number, $ 0.66 {\rm fm}$. Two points are worth stressing
here. Firstly, it is clear that the finite cut-off corrections go in
the wrong direction. On the other hand, one should not expect a
perfect agreement with the experimental number, since one expects a
sizeable contribution from the pion loops ($1/N_c$ corrections in the
model), as suggested by ChPT~\cite{GL84}. There is a way one can get
rid of the cut-off corrections; one can simply split the charge
contribution at $q^2=0$, which should be one, and compute the
difference using a unregularized quark loop. This is equivalent to
write down a once subtracted dispersion relation for $F_\pi^{\rm em}
(q^2)$. For the unregularized quark loop only one has the asymptotic
behavior for $q^2 \to -\infty $, 
\begin{eqnarray}
F_\pi^{\rm em} (q^2) \sim \log ( -q^2 /M^2),
\end{eqnarray} 
without power corrections. For the PV regularized NJL model, at high
Euclidean momentum one has instead a cut-off independent relation,
which is a pure power correction~\footnote{For a Euclidean O(4)
cut-off one gets a logarithmic behavior $\log ( Q^2 / \Lambda^2
)/Q^2$~\cite{BH88}. },
\begin{eqnarray}
q^2 F_\pi^{\rm em} (q^2) \to -\frac{M \langle \bar q q \rangle }{2 
f_\pi^2 } \sim 0.34 {\rm GeV}^2
\end{eqnarray} 
for $M=300 {\rm MeV} $ and the PV method, Eq.~(\ref{eq:PV2}). Up to
the measured momentum transfers, $-2 {\rm GeV}^2 < q^2 <  -6 {\rm
GeV}^2 $ this value is very similar to the experimental average number
$0.38 \pm 0.04$~\cite{FF}. However, it is in principle not clear if
this number should directly be compared to experiment, without taking
into account QCD radiative corrections. Another point of view is to
reject the validity of the model to such high energies, matching
instead to the known QCD result at LO~\cite{FJ79} (see also
Ref.~\cite{BL80},
\begin{eqnarray}
q^2 F_\pi^{\rm em} (q^2) \to 16 \pi \alpha (-q^2) f_\pi^2 
\label{eq:pff_qcd} 
\end{eqnarray} 
which for $q^2=- 4 {\rm GeV}^2 $ yields, $ 0.13 $, a too small value
as compared to the experimental one~\footnote{Actually, this
expression can be obtained from 
\begin{eqnarray} 
-Q^2 F_\pi^{\rm em} (-Q^2) \to 16 \pi \alpha (Q^2) f_\pi^2 \left(
\int_0^1 \frac{\varphi_\pi (x,Q)}{6 x(1-x)} \right)^2
\label{eq:pion_ffpda} 
\end{eqnarray} 
with $\varphi_\pi (x,\mu) $ the pion distribution amplitude at the
scale $\mu$, when the asymptotic wave function, $\varphi_\pi
(x,\infty) = 6 x(1-x)$ is substituted~\cite{BL80} (see also
below).}. We may identify both coefficients at the model scale,
$\mu_0$, yielding the result
\begin{eqnarray}
\frac{\alpha (\mu_0)}{2\pi} = -\frac{M_{\mu_0} \langle \bar q q
\rangle_{\mu_0} }{64 \pi^2 f_\pi^4 }
\label{eq:mu0_piff}
\end{eqnarray} 
In this relation we have stressed the fact that also the constituent
quark mass is fixed at the model scale, $\mu_0$. Actually, we may use
relations (\ref{eq:mu0_piff}) and (\ref{eq:mu0_dis}) to get,
\begin{eqnarray}
M_{\mu_0}= 300 \pm 80 {\rm MeV}
\label{eq:m_m0}
\end{eqnarray}
Although this value should be considered a crude estimate it provides
a {\it reasonable} value for the constituent quark mass at the model
scale. Conversely, if we assume $M=300 {\rm MeV}$ then $ \alpha(\mu_0)
/(2\pi) \sim 0.5 \pm 0.2 $, a compatible value with the momentum
fraction estimate. 
 
\section{Pion Distribution Functions} 

\subsection{Deep Inelastic Scattering} 

Inclusive lepton-hadron scattering is described in terms of the
hadronic tensor, $W_{\mu\nu} (p,q)$, and can be obtained from the
imaginary part of the forward Compton amplitude for virtual photons as
follows~\cite{Ja85}
\begin{eqnarray} 
 W_{\mu\nu} (p,q) &=& {1\over 2\pi} {\rm Im} \, T_{\mu\nu} (p,q) \\ =
W_1 ( q^2 , p \cdot q ) \left( - g_{\mu\nu} + {q_\mu q_\nu \over q^2
}\right) &+& {W_2 ( q^2 , p\cdot q )\over m_\pi^2 } \left( p_\mu - {p
\cdot q \over q^2 } q_\mu \right) \left( p_\nu - {p \cdot q \over q^2
} q_\nu \right), \nonumber
\end{eqnarray} 
where
\begin{eqnarray} 
T_{\mu\nu} ( p,q) = i \int d^4 x e^{i q \cdot x }
\langle \pi(p)|T \Bigr\{ J_\mu^{\rm em} (x) J_\nu^{\rm em}
(0)\Bigr\}|\pi(p) \rangle.
\label{virtual-compton}
\end{eqnarray} 
\begin{figure}[tbp]
\begin{center}
\epsfig{figure=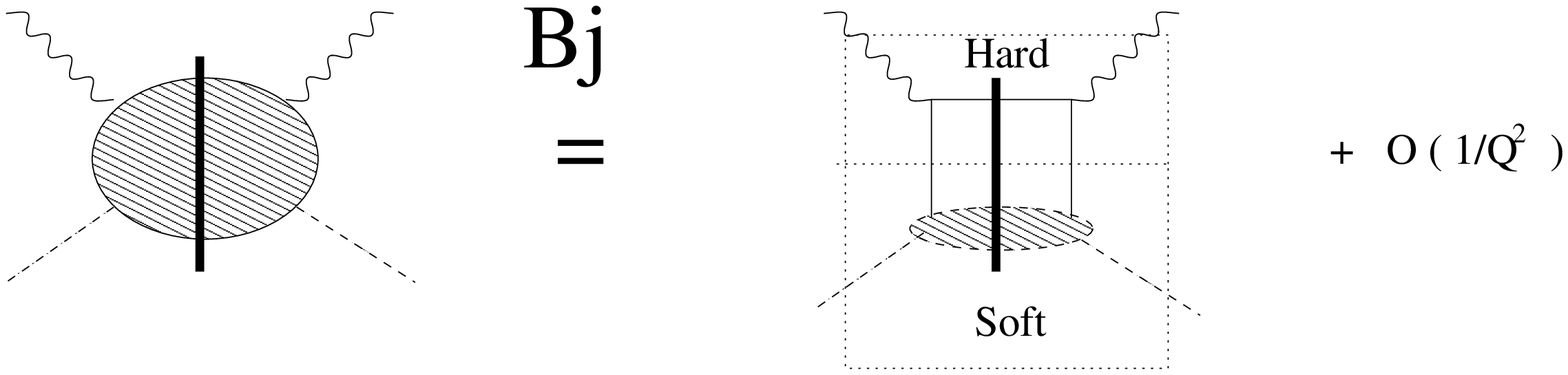,height=3cm,width=10cm}
\end{center} 
\caption{The Compton amplitude in the Bjorken limit and its relation
with the quark-target scattering in the parton model. The high and low
energy contributions are also explicitly displayed.}
\label{fig:quark-target}
\end{figure}
Here, $m_P^2 = p^2 $ is the mass of the pseudoscalar meson and $q$ is
the momentum of the virtual photon. In the Bjorken limit, one
obtains~\cite{Ja85} $ ( x= -q^2 / (2 p\cdot q ) )$,
\begin{eqnarray} 
&& W_1 (x, Q^2) \to F_1 (x) \\ 
&& W_2 (x, Q^2) \to F_2 (x) 
\end{eqnarray} 
where the Callan-Gross relation $ F_2 (x) = 2 x F_1 (x) $ is
fulfilled, as a consequence of the spin 1/2 nature of quarks. The
structure function is given by
\begin{eqnarray} 
F_1 (x) = {1\over 2} \sum_{i=u,d} e_i^2 [ \bar q_i (x) + q_i (x) ].
\end{eqnarray} 
where $e_u=2/3$ and $e_d=-1/3$ are the quark charges and $q_i (x)$ and
$\bar q_i(x) $ are the momentum fraction distribution of the different
quark species. 

Introducing light-cone (LC) coordinates $x=(x^+,x^-,
\vec x_\perp)$ with $x^{\pm}=x^0 \pm x^3 $ the analysis of
Ref.~\cite{Ja85} yields
\begin{eqnarray}
F_1 (x) = -{i\over 4 \pi} \int { d k^- d^2 \vec k_\perp \over
(2\pi)^3 } {\rm tr} \left[ \hat Q^2 \gamma^+ \chi(p,k) \right]
\Big|_{k^+ = p^+= m_\pi x}
\label{quark-target} 
\end{eqnarray} 
where $\hat Q={\rm diag}( e_u, e_d) $ is the charge operator and the
forward quark-target scattering amplitude is defined
\begin{eqnarray}
\chi(p,k) = -i \int d^4 \xi \, e^{i \xi \cdot k} \langle p | T \{ q(\xi) \bar
q(0) \} | p \rangle 
\label{q-t} 
\end{eqnarray} 
$\chi(p,k)$ corresponds to the unamputated
vertex. Eq.(\ref{quark-target}) holds under the assumption of scaling
and finiteness in the Bjorken limit. 

Logarithmic scaling violations due to perturbative QCD radiative
corrections~\cite{AP77}, relate in a linear fashion the leading twist
contribution to structure functions at a given reference scale,
$\mu_0$, to the scale of interest, $\mu$, (in DIS one takes $Q^2
=\mu^2$)
\begin{eqnarray} 
F_1 (x, \mu ) = U(\mu , \mu_0 ) F_1 (x,\mu_0 ).
\label{evol} 
\end{eqnarray} 
where Eq.~(\ref{eq:cocycle}) is satisfied for the linear matrix
integral operator $U$. Asymptotically, QCD predicts~\cite{AP77}  
\begin{eqnarray}
F_2 (x, Q^2 ) = 2 x F_1 (x, Q^2) \to \frac{\sum_i e_i^2 }{1 + 2(N_c^2
-1)/N_F } \, \delta (x)
\label{eq:Finfty}
\end{eqnarray}  
Actually, the pion structure functions cannot be measured directly
since there are no pion targets. Instead, the parton structure of the
pion has been analyzed \cite{SMRS92} and a simple parameterization at
$Q^2 = 4 {\rm GeV}^2$ has been given. The valence quark distributions
extracted in Ref.~\cite{SMRS92} from Drell-Yan experiments \cite{Co89}
seem well determined, whereas the gluon distribution as obtained from
the analysis of prompt-photon emission data \cite{Au89} is less well
determined. A recent analysis \cite{kla01} of the ZEUS di-jet data
seems to favor the gluon distributions of Ref.~\cite{SMRS92}. This is
why, in determining the low-energy scale of our model, we use the
valence momentum fraction of $47 \%$ found in Ref.~\cite{SMRS92} at $4
{\rm GeV}^2$ instead of $40 \% $ found Ref.~\cite{GRS98,GRS99} using
some additional assumptions~\cite{AP96}.

A result found long ago~\cite{bf73} and usually adopted in DIS
calculations are the so-called counting rules. They establish that
there is a relation between the asymptotic behavior of the
electromagnetic pion form factor for large Euclidean momenta and the
structure function behavior as $x \to 1 $, namely
\begin{eqnarray}
F_\pi^{\rm em} (q^2 ) \sim 1/ (q^2)^n \qquad F_1 (x) \sim (1-x)^{2n-1}   
\label{eq:counting} 
\end{eqnarray} 
In QCD $ F_\pi^{\rm em} (q^2 ) $ has a logarithmic radiative
correction, Eq.~(\ref{eq:pff_qcd}) so we may take $n \sim 1$, which
implies $ F(x) \sim 1-x $. Although it is not clear at what scale
$\mu$ the counting rules are valid, recent investigations based on
quark-hadron duality confirm the exponent $n \sim 1$ for $F(x)$ at $
\mu \sim 2 {\rm GeV} $~\cite{Me02}.

\subsection{Spectator model and quark loop calculations} 

The spectator model is a very simple model where one can illustrate
the source of the problems in this kind of calculations and the
relation of gauge invariance and normalization. Actually, in the case
of the pion such a model corresponds to a unregularized NJL
calculation. The spectator model consists of replacing the full sum of
intermediate states in the soft piece of the Compton amplitude by a
spectator particle, which for the pion has quark quantum
numbers~\cite{JM97}. All is required to do the calculation is to know
the $\pi q\bar q$-coupling. We take a pseudoscalar coupling as follows
\begin{eqnarray}
\Gamma_{\pi qq}^a = - \gamma_5 \tau^a g_{\pi qq}
\end{eqnarray} 
At the level of the Compton scattering amplitude it corresponds to use
the well-known hand-bag diagram. Straightforward calculation of the
unregularized structure function yields in the Bjorken limit
\begin{eqnarray}
F_1 (x,Q^2 ) &\to & \frac12 (e_u^2 + e_d^2 ) \frac{4 N_c g_{\pi
qq}^2}{(4\pi)^2 } \nonumber \\ & \times & \left\{ -\log
\left[\frac{M^2 -x(1-x) m_\pi^2 }{Q^2} \right] + \frac{m_\pi^2 x
(1-x)}{M^2 - x(1-x) m_\pi^2} \right\}
\label{eq:F_unreg} 
\end{eqnarray} 
which is a scaling violation quite different from those found in QCD.
Actually, we get $F_2 (x,\infty)= \infty $ instead of
Eq.~(\ref{eq:Finfty}). Scaling can be restored by attaching a form
factor to the $\pi qq $ vertex or putting a $O(4)$ cut-off in the
loop. No solution is really free of problems, since either gauge
invariance is violated or extra singularities are introduced.

At the level of the quark-target scattering amplitude the spectator
picture corresponds to consider an intermediate constituent quark
state. The result for a pseudoscalar coupling is
\begin{eqnarray}
F_1 (x)= \frac12 (e_u^2 + e_d^2 ) \frac{4 N_c g_{\pi qq}^2}{4\pi} \int
\frac{d^2 k_\perp}{(2\pi)^2} \frac{k_\perp^2+M^2 }{[ k_\perp^2 + M^2 -
x(1-x) m_\pi^2]^2 }
\label{eq:qt-spec} 
\end{eqnarray} 
However, we see that the integral in $k_\perp$ is actually
logarithmically divergent. This only reflects the fact that for the
unregularized quark loop the separation between soft and hard
processes involved in the Bjorken limit takes place at $k_\perp^2 \sim
Q^2$. Obviously a form factor for the vertex or a transverse cut-off,
$|k_\perp| \le \Lambda $ (both $Q^2$ independent), may be introduced
yielding a finite result.

Thus, we learn in this particular model that the connection between
the Compton amplitude and the quark-target scattering formulas is only
justified provided everything is finite. So, some high energy
suppresion should be applied, with the help of a regularization. The
problem is that we expect to do it in a gauge invariant manner, since
after all we are dealing with an electromagnetic process.

\begin{figure}[tbp]
\begin{center}
\epsfig{figure=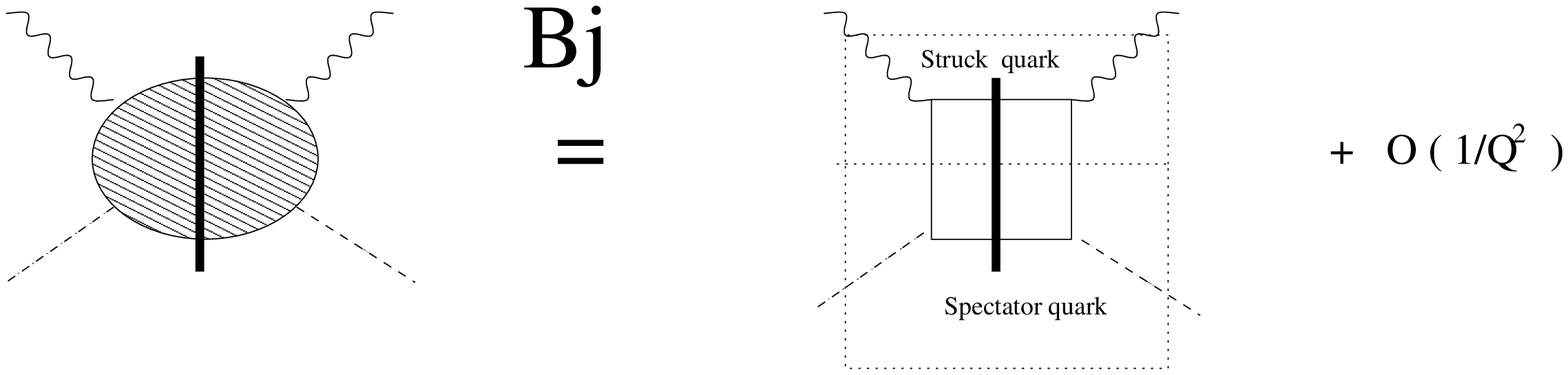,height=3cm,width=10cm}
\end{center} 
\caption{Spectator model. The sum over intermediate states in the soft
hadronic contribution is replaced by a single quark state.}
\label{fig:spectator}
\end{figure}

\subsection{PDF Calculation from the Compton amplitude} 

\begin{figure}[tbp]
\begin{center}
\epsfig{figure=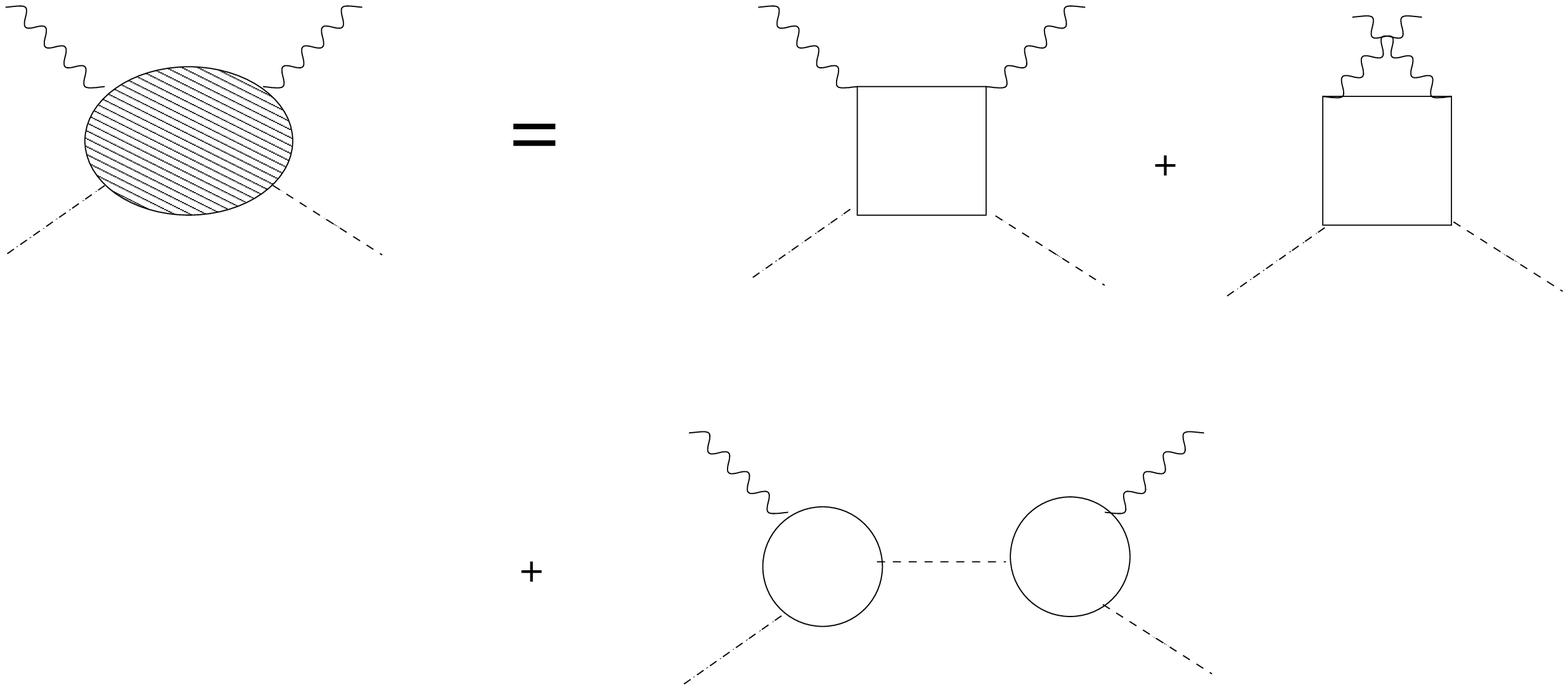,height=4cm,width=10cm}
\end{center} 
\caption{\footnotesize Gauge invariant Compton amplitude in the NJL
model. The closed loops are Pauli-Villars regularized. Wavy lines are
photons, solid lines are constituent quarks and dashed lines composite
pions.}\label{fig:compton_njl}
\end{figure}

In the PV regularized NJL model, the gauge invariant forward virtual
Compton scattering amplitude is given the sum of the hand-bag diagram,
the crossed contribution and the $ \gamma \pi \to \pi* \to \pi \gamma$
involving the off-shell electromagnetic pion form factor (see
Fig.~(\ref{fig:compton_njl})). Actually, it can be shown that in the
Bjorken limit only the hand-bag contribution survives, since the
crossed and pion off-shell contributions are higher twist. The result
has been obtained in Refs.~\cite{DR95,WRG99} and turns out to coincide
with applying PV regularization, Eq.(\ref{eq:PV}), to Eq.~(
\ref{eq:F_unreg}). The result is
\begin{eqnarray}
F_1 (x)= \frac12 (e_u^2 + e_d^2 ) 4 N_c g_{\pi qq}^2 \frac{ d }{d
m_\pi^2 } \left[ m_\pi^2 F (m_\pi^2, x ) \right]
\end{eqnarray} 
where the function $ F(x,p^2) $ is given by Eq.~(\ref{eq:Fpx}). Taking
$\pi^+$ for definiteness, one gets in the chiral limit
\begin{eqnarray}
u_{\pi^+ } (x ) = \bar d_{\pi^+} (x ) = 1 \qquad , \,  0 < x < 1 \, .  
\label{result0} 
\end{eqnarray} 
This means the following result for the valence-quark distribution function
\begin{eqnarray}
V(x)= 2 \qquad \, 0 < x < 1  \, . 
\label{result} 
\end{eqnarray} 
As we see the result is independent of the PV regulators and has
proper support, normalization and fulfills the momentum sum
rule. Moreover, Eq.(\ref{result}) is consistent with chiral symmetry,
in the sense that it has been obtained taking explicitly into account
the boson Goldstone nature of the pion. The result has also been
obtained within the NJL model imposing a transverse cut-off
\cite{BH99} upon the quark-target amplitude (\ref{eq:qt-spec}).

Eq.(\ref{result}) disagrees with other quark loop calculations. If
Eq.(\ref{virtual-compton}) is used with a four-dimensional cut-off
\cite{SS93} or Eq.(\ref{quark-target}) is used with Lepage-Brodsky
regularization \cite{BH99} in the NJL model, different shapes for the
quark distributions are obtained. The null-plane \cite{FM94}, NJL
model \cite{SS93}, spectator model \cite{JM97} and the recent quark
loop \cite{BC02} calculations also produce different results. In all
cases, the use of momentum dependent form factors or non-gauge
invariant regularization make the connection between
Eq.(\ref{virtual-compton}) and Eq.(\ref{quark-target}) doubtful and,
furthermore, spoil normalization. The results based on a quark loop
with momentum dependent quark masses \cite{DT00,HRS01} seem to produce
a non-constant distribution.

\subsection{PDF Calculation in Light-Cone Coordinates} 

Perhaps the simplest way to obtain PDF for the pion in the NJL model
is by using the pion Bethe--Salpeter propagator. The probability
distribution is defined through the pion expectation value of the of
number operator corresponding to a quark of a given species.  Taking
into account the proper normalization factors one gets at the pion
pole the following expression for the probability distribution of
finding a valence quark with momentum fraction $x$ and transverse
momentum $\vec k_\perp $
\begin{eqnarray}
\frac12 V_\pi (x,k_\perp ) = \int d k^-  
D(p,k) \Big|_{k^+ = m_\pi x, p^2=m_\pi} 
\label{eq:pdf} 
\end{eqnarray}  
where the Bethe--Salpeter normalization in the Pauli--Villars
regularization scheme is
\begin{equation}
D (p,k) = 4N_c {\rm i} g_{\pi qq}^2 {d\over d p^2} \Big\{ p^2 \sum_i
c_i {1 \over k^2 - M^2 - \Lambda_i^2 + {\rm i}\epsilon}\, {1\over
(k-p)^2 - M^2 - \Lambda_i^2 + {\rm i}\epsilon} \Big\}
\Big|_{p^2=m_\pi^2 } \, .
\end{equation}
Introducing light--cone (LC) variables
\begin{eqnarray}
k^+ = k^0 + k^3 \, , \quad k^- = k^0 - k^3  \, , \quad d^4 k =
{1\over 2} dk^+ dk^- d^2 k_{\perp} \, .
\end{eqnarray}
Employing the definition (\ref{eq:pdf}) one finds 
\begin{eqnarray}
\frac12 V_{\pi } (x, \vec k_\perp ) &=& -\frac{ 2 i N_c g_{\pi qq}^2 }{
f_\pi} \int \frac{dk^+ dk^- }{(2\pi)^4 } \frac1 { x(1-x) }\times
\nonumber \\ && \frac{d}{dm_\pi^2} \left\{ \sum_j  c_j  \frac{\delta
\left( k^+ - x p^+ \right)} {k^- - m_\pi - \frac{ \vec k_\perp^2 + M^2
+ \Lambda_j^2 + i 0^+ }{m_\pi (1-x) } } \, \frac1{k^- - \frac{ \vec
k_\perp^2 + M^2 + \Lambda_j^2 + i 0^+}{m_\pi x } } \right\} \nonumber,
\label{eq:LC_int} 
\end{eqnarray}
where the location of poles in the $k^-$ variable has been explicitly
displayed.  For $x>1$ or $x<0$ both poles are above and below the real
axis respectively, and hence the integral vanishes in either case. For
$ 0 < x < 1 $ the integral yields
\begin{eqnarray}
\frac12 V_\pi (x, k_\perp ) = {4N_c g_{\pi qq}^2 \over 16 \pi^3 }
{d\over d m_\pi^2} \left\{ m_\pi^2 \sum_i c_i {1\over k_\perp^2 + m^2
+ \Lambda_i^2 -m_\pi^2 x(1-x) } \right\} \, .
\end{eqnarray}
Notice that, due to the Pauli--Villars subtractions, we have the
asymptotic behavior in the transverse momentum $k_\perp$,
\begin{eqnarray}
\frac12 V_\pi (x, k_\perp ) \sim {4N_c g_{\pi qq}^2 \over 16\pi^3}
{\sum_i c_i \Lambda_i^4 \over k_\perp^6} \, .
\end{eqnarray}
This guarantees the convergence of the $k_\perp$ integral without
introducing a transverse cut--off. Thus, the Pauli--Villars regulators
automatically provide a (gauge-invariant) form of a transverse
cut-off.  Integrating the transverse momentum we get the PDF
\begin{eqnarray}
\frac12 V_\pi (x) &=& \int d^2 k_\perp \frac12 V_\pi (x ,\vec k_\perp
) \\ &=& 4 N_c g_{\pi qq}^2 \frac{ d }{d m_\pi^2 } \left[ m_\pi^2 F
(m_\pi^2, x ) \right] 
\end{eqnarray}
which is the result found in Ref.~\cite{DR95} (see also
Eq.~(\ref{eq:qt-spec}) above).  In the chiral limit, $m_\pi = 0$, one
can use the Goldberger-Treiman relation for the constituent quarks, $
g_{\pi qq} f_\pi = M$. Then $f_\pi^2 = 4N_c M^2 F(0)$, which gives the
very simple formulas
\begin{eqnarray} 
\frac12 V_\pi ( x, \vec k_\perp ) &=& 
 \frac{4 N_c M^2}{16\pi^3 f_\pi^2} \sum_i
c_i \frac1{k_\perp^2 + \Lambda_i^2 + M^2 } ,
\label{eq:td0} 
\\ \frac12 V_\pi (x) &=& 1 .\label{eq:pdf0} 
\end{eqnarray} 
In the chiral limit $V_\pi ( x, \vec k_\perp ) $ becomes trivially
factorizable, since it is independent of $x$. A remarkable feature is
that the last relation, Eq.~(\ref{eq:pdf0}), is independent of the PV
regulators.

The light--cone interpretation has been
pursued (see {\it e.g.} \cite{FM94} and references therein) and more
recently \cite{He00,BH99} within a LC quantization. In these cases
transverse cut--off's were introduced, {\it a posteriori}. As we have
shown above this is not necessary in the PV regularization.

\subsection{QCD evolution} 

\begin{figure}[tbp]
\begin{center}\epsfig{figure=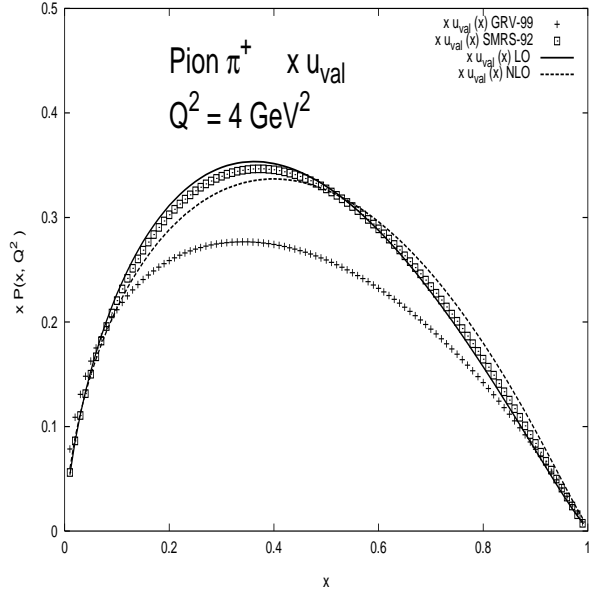,height=8cm,width=8cm}
\end{center} 
\caption{Valence distributions in the pion at $Q^4 = 4 {\rm GeV}^2 $
at LO~\cite{DR95} and NLO~\cite{DR02} compared with phenomenological
analysis for the pion SMRS92 \cite{SMRS92}, GRV99 \cite{GRS99}.  We
take $ \langle x V \rangle_\pi = 0.47$ at $ Q^2 = 4 {\rm GeV}^2
$~\cite{SMRS92}.}
\label{fig:pionval}
\end{figure}

Due to radiative corrections, parton distribution functions evolve
logarithmically with scale through the DGLAP
equations~\cite{AP77}. Non-singlet (or valence) distribution functions
are easily evolved in terms of their Mellin moments. For $\pi^+ = u
\bar d $ one defines
\begin{eqnarray}
V(x) = u(x) - \bar u (x)  + \bar d (x) - d (x)  
\end{eqnarray} 
and takes (we assume chiral limit for simplicity)~\footnote{For non
experts it may sound unnatural to take the high energy, $Q^2\to \infty
$, limit in a model and use it as initial condition for QCD evolution
at a low scale $\mu_0$. The point is that if the quark model scales,
the asymptotic behavior can be separated according to increasing power
corrections in $1/Q^2$ in a twist expansion. The anomalous dimensions
relevant for QCD evolution of structure functions are only known for
the lowest orders of such an expansion, but not for the full structure
function. Using Eq.~(\ref{eq:vpimu0}) as initial condition
automatically complies with the asymptotic result of
Eq.~(\ref{eq:Finfty}).}  \begin{eqnarray} \frac12 V_\pi (x, \mu_0) = 1
\label{eq:vpimu0} 
\end{eqnarray} 
At leading order, evolution reads
\begin{eqnarray}
V_n (\mu) \equiv \int_0^1 dx x^{n} V(x,\mu) = \left(
\frac{\alpha(\mu)} {\alpha(\mu_0) } \right)^{\gamma_{n}^{\rm NS} / 2
\beta_0 } \int_0^1 dx x^{n} V(x,\mu_0)
\label{eq:evol_val} 
\end{eqnarray} 
where the anomalous dimension is 
\begin{eqnarray}
\gamma_n^{\rm NS} &=& -2 C_F \left[ 3 + \frac{2}{(n+1)(n+2)}- 4
\sum_{k=1}^{n+1} \frac1k \right],
\label{eq:anom_dim} 
\end{eqnarray}
with $C_F = 4/3$. Taking $n$ as a complex number, which also requires
an analytical continuation of both $V_n (\mu_0) $ and $\gamma_n^{\rm
NS}$, Eq.~(\ref{eq:evol_val}) can be inverted using
\begin{equation}
V (x, \mu )=\int_{c-i\infty}^{c+i\infty}{{\rm d} n\over 2\pi{\rm i}}
x^{-n} V_n (\mu)
\label{eq:inv_mellin} 
\end{equation}
where $c$ has to be chosen as to leave all the singularities on the
left hand side of the contour. The result for the pion valence PDF
found in Ref.~\cite{DR95} at LO and \cite{DR02} at NLO is displayed in
Fig.~(\ref{fig:pionval}) at $\mu^2 = Q^4 = 4 {\rm GeV}^2 $ compared
with phenomenological analysis for the pion \cite{SMRS92}. As one can
see, the agreement is quite impressive. Despite the fact that $\alpha
(\mu_0) / (2\pi)  \sim 0.3 $ the differences between LO and NLO turn
out to be small. 

It is interesting to analyze these results on the light of the
counting rules, Eq.~(\ref{eq:counting}), of Ref.~\cite{bf73}, as was
already done in Ref.~\cite{Ru98b}. Using the inverse Mellin formula,
Eq.~(\ref{eq:inv_mellin}), it can be shown that, if $ V (x,\mu_0) \to
C (1-x)^N $ then
\begin{eqnarray} 
V (x,\mu) \to C (1-x)^{N - \frac{4 C_F }{\beta_0} \ln \frac{\alpha(\mu)}{
\alpha(\mu_0) }} \qquad x\to 1 
\end{eqnarray} 
The value of the additional contribution to the exponent is a weakly
depending function for $\mu > 1 {\rm GeV}$. For the pion we find, using
Eq.~(\ref{eq:mu0_dis}), 
\begin{eqnarray} 
V (x, 2 {\rm GeV} ) {\to} 2 (1-x)^{1.1 \pm 0.1}  \qquad x \to 1 
\end{eqnarray} 
which is consistent with the counting rules and the phenomenological
analysis~\cite{SMRS92}, {\it after evolution}. A more detailed account
of the sea and gluon distribution functions at LO and NLO and also for
$K$ and $\eta$ mesons can be found in Ref.~\cite{DR02}.

\section{Pion distribution amplitude} 

\subsection{Pion Transition Form Factor} 

The matrix element for the $ \gamma^* \to \pi^0 \gamma $ transition
form factor is defined
\begin{eqnarray}
\Gamma^{\mu \nu}_{\pi^0 , \gamma ,\gamma } (p,q_1,q_2)  &=& 
\epsilon_{\mu\nu \alpha \beta}q_1^\alpha q_2^\beta F_{\pi \gamma \gamma}
(p,q_1,q_2)
\end{eqnarray} 
where $F_{\pi \gamma \gamma} (p,q_1,q_2)$ is the $\pi^0$
electromagnetic transition form factor~\cite{BL80}. Two QCD results
apply for this transition form factor. In the chiral limit and for
on-shell pions, $p^2=0$, and photons, $q_1^2 =q_2^2 = 0$,
corresponding to the $\pi^0 \to 2 \gamma$ decay one gets the
normalization condition 
\begin{eqnarray}
F_{\pi \gamma \gamma} (0,0,0) = \frac{1}{4 \pi^2 f_\pi}
\label{eq:anom} 
\end{eqnarray} 
which is the standard result expected from the chiral anomaly.  The
definition of the leading twist contribution to the pion distribution
amplitude (PDA), $\varphi_\pi (x)$, from the transition form factor is
\begin{eqnarray}
 \lim_{Q^2 \to \infty } & Q^2 & F_{\pi \gamma \gamma} ( p,q_1,q_2 )
\Big|_{q_1^2 = -\frac{(1+\omega)}2 Q^2 , q_2^2 = -\frac{(1-\omega)}2
Q^2 } \nonumber \\ &=& \frac{4 f_\pi}{N_c} \int_0^1 dx
\frac{\varphi_\pi (x) }{1-\omega (2x-1)}
\label{eq:tranpda}
\end{eqnarray} 
where $ \omega \equiv (q_1^2 - q_2^2)/(q_1^2+ q_2^2)$ is the photon
asymmetry, $ -1 < \omega < +1 $, which is kept finite when taking the
limit $Q^2 \to \infty $. It can be shown that in the parton model the
PDA can be computed by 
\begin{eqnarray}
\varphi_\pi (x) = \int d^2 k_\perp \Psi_{\pi } (x, k_\perp), 
\end{eqnarray}  
where the pion light-cone wave function (pseudoscalar component) is
defined as the low-energy matrix element~\cite{LB81,St99,Br01} 
\begin{eqnarray}
&&\Psi_{\pi } (x, \vec k_\perp ) = -\frac{{i} \sqrt{2} }{4\pi
f_\pi} \int d \xi^- d^2 \xi_\perp e^{{i} (2x-1) \xi^- p^+ -
\xi_\perp \cdot k_\perp } \nonumber 
\times \\ && \langle \pi^+ (p) | \bar u (\xi^- ,
\xi_\perp) \gamma^+ \gamma_5 d(0) | 0 \rangle .
\label{eq:pda_def}
\end{eqnarray} 
where $p^\pm = m_\pi $ and $\vec p_\perp= \vec 0_\perp $. An important
relation found in Ref.~\cite{BL80} reads
\begin{eqnarray}
f_\pi \int_0^1 dx \Psi_\pi ( x, 0_\perp ) = F_{\pi \gamma \gamma} (0,0,0)
\label{eq:lc_anom} 
\end{eqnarray} 
Radiative logarithmic corrections to the pion distribution amplitude
(PDA) can be easily implemented through the QCD evolution
equations~\cite{BL80}, which yield for $Q^2 \to \infty$ the asymptotic
wave function of the form 
\begin{eqnarray} 
\varphi_\pi(x,\infty) = 6 x(1-x). 
\label{eq:phi_asymp}
\end{eqnarray} 
Moreover, the pion transition form factor has been recently measured
by the CLEO collaboration~\cite{CLEO98} and a theoretical analysis of
PDA based on these data and light-cone sum rules has been
undertaken~\cite{SY00}, showing that at $Q=2.4$~GeV PDA is neither
asymptotic, nor possesses the structure proposed in early
works~\cite{CZ84}.

The pion distribution amplitude has been evaluated with QCD sum
rules~\cite{BJ97}, in standard~\cite{DP00} (only the second
$\xi$-moment) and transverse lattice approaches~\cite{Da01,BS01,BD02},
and in chiral quark
models~\cite{PP97,PP99,ADT00,He00,He01,PR01,ADT01,DVY02,Do02}. In
chiral quark models the results are not always compatible to each
other, and even their interpretation has not always been the
same. While in same cases there are problems with chiral symmetry and
proper normalization~\cite{PP97,PP99,PR01}, in other cases
\cite{ADT00,He00,He01,PR01,ADT01,DVY02,Do02} it is not clear how to
associate the scale at which the model is defined, necessary to define
the starting point for the QCD evolution. The fact that several
calculations~\cite{PP97,PP99,He00,He01,PR01,ADT01,DVY02} produce a PDA
strongly resembling the asymptotic form suggests that their working
scale is already large, and the subsequent QCD evolution becomes
unnecessary, or numerically insignificant.  This also tacitly assumes
that these models already incorporate the QCD radiative corrections.

\subsection{PDA calculation for the unregularized quark loop} 

Since $\gamma^* \to \pi^0 \gamma $ is an abnormal parity process, the
standard procedure in the NJL model is not to regularize it because
this is the only way to preserve the anomaly (See also
Ref.~\cite{BH88,SR95})). Straightforward calculation yields the result
\begin{eqnarray}
F_{\pi \gamma \gamma} (p,q_1,q_2) = 8 g_{\pi qq} M  \frac1{\rm i}\int
 {d^4 k \over (2\pi)^4}
 \frac1{k^2-M^2}\frac1{(k-q_1)^2-M^2}\frac1{(k-q_2)^2-M^2}
\end{eqnarray} 
In the chiral limit $ g_{\pi qq} f_\pi \to M $
and for on-shell pions, $p^2=0$, and photons, $q_1^2 =q_2^2 = 0$,
corresponding to the $\pi^0 \to 2 \gamma$ decay one gets
\begin{eqnarray}
F_{\pi \gamma \gamma} (0,0,0) = \frac{8 M^2 }{f_\pi} \frac1{\rm i}\int
{d^4 k \over (2\pi)^4} \frac1{(k^2-M^2)^3} = \frac{1}{4 \pi^2 f_\pi}
\end{eqnarray} 
which agrees with the chiral anomaly expectations,
Eq.~(\ref{eq:anom}). This result is the main motivation for not
introducing an explicit cut-off in the abnormal parity processes. In
order to compute the the high momentum behavior, we use the Feynman
trick in the two propagators containing $q_1$ and $q_2$, and shift the
integrating variable. For on-shell massless $\pi^0$, $p^2=0$, we get
\begin{eqnarray}
F_{\gamma^* \gamma^* \pi} &=& \frac{ 8 M^2}{f_\pi} \int {d^4 k \over
(2\pi)^4} \frac1{(k^2-M^2)^2} \nonumber \\ && \times \frac1{\rm i}\int_0^1 dx
\frac1{(k+x q_1 + (1-x)q_2 )^2-M^2}\,
\end{eqnarray}
The $Q^2 \to \infty $ limit can be undertaken yielding (see
Eq.~(\ref{eq:tranpda}))
\begin{eqnarray}
\varphi_\pi (x, Q^2 ) \to -\frac{4 N_c g_{\pi qq} M}{(4\pi)^2} \log
\left[\frac{M^2 -x(1-x) m_\pi^2 }{Q^2} \right] \, , 
\label{eq:phi_unreg} 
\end{eqnarray} 
where a finite pion mass has been reinserted.  Thus, the prescription
not to regularize the abnormal parity vertex does not agree with the
factorization result found in a parton model approach. Instead, there
appear scaling violations, which do not correspond to those expected
from QCD evolution, contradicting the asymptotic result,
Eq.~(\ref{eq:phi_asymp}). This problem is of a similar nature as the
one found in the discussion of the spectator model after
Eq.~(\ref{eq:F_unreg}), and similar remarks concerning gauge
invariance apply here.

\begin{figure}[tbp]
\begin{center}
\epsfig{figure=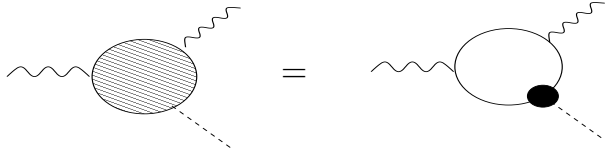,height=2cm,width=8cm}
\end{center} 
\caption{\footnotesize Transition form factor in the NJL model.  The
closed loop is Pauli-Villars regularized. Wavy lines are photons,
solid lines are constituent quarks and dashed lines composite
pions.}\label{fig:trans}
\end{figure}

\subsection{PDA calculation from the transition form factor} 

To achieve factorization, it is necessary to effectively cut-off the
transverse momentum, i.e., we have to be able to make $Q^2$ larger
than any scale in the loop. This is not consistent with integrating in
$k$ up to infinity\footnote{If we formally take the limit we get an
expression looking like Eq.~(\ref{eq:tranpda}) with, $\varphi_\pi (x)=
1$ but with the unregularized form of $f_\pi^2 $ ( see
Eq.~(\ref{eq:fpi}). To achieve Eq.(\ref{eq:tranpda}) with the
regularized definition of $f_\pi^2 $ some regularization must be
introduced.}. If we introduce PV regulators~\footnote{This
regularization guarantees that the momentum routing is irrelevant. The
Euclidean $O(4)$ cut-off of Ref.~\cite{ADT00,DVY02} requires a very
special momentum routing {\it after} regularization; a different
momentum routing with the same cut-off would produce finite cut-off
contributions to their results.}, we get
\begin{eqnarray}
Q^2 F_{\gamma^* \gamma^* \pi} (p,q_1,q_2) |_{\rm reg.} & \to & \frac{
16 M^2}{f_\pi} \sum_i c_i \frac1{\rm i}\int {d^4 k \over (2\pi)^4}
\frac1{(k^2-M^2-\Lambda_i^2 )^2} \nonumber \\ & \times & \int_0^1 dx
\frac1{1-\omega (2x-1)}
\end{eqnarray}
which, using Eq.~(\ref{eq:fpi}) and Eq.~(\ref{eq:tranpda}) corresponds
to take (after inclusion of a finite $m_\pi$ ) 
\begin{eqnarray} 
\varphi_{\pi} (x)= 4 N_c M \frac{g_{\pi qq}}{f_\pi} F(m_\pi^2 , x ) , 
\label{eq:pda}
\end{eqnarray}
where $F(p^2,x)$ is given in Eq.~(\ref{eq:Fpx}). $\varphi_\pi(x)$
depends only on $x$ and is properly normalized (see
Eq.~(\ref{eq:fpi})). In the chiral limit one gets
\begin{eqnarray} 
\varphi_{\pi} (x)= 1 \qquad 0 < x < 1 \, . 
\label{eq:pd0}
\end{eqnarray}
independently on the regulators. However, if we PV regularize the
vertex there is a violation of the anomaly, since
\begin{eqnarray}
F_{\pi \gamma \gamma} (0,0,0) |_{\rm reg} = \frac{1}{4 \pi^2 f_\pi}
\sum_i c_i \frac{M^2}{M^2+\Lambda_i} 
\end{eqnarray} 
For typical values of the parameters one finds a $25 \% $ reduction in
the amplitude which means a $40 \%$ reduction in the width for $\pi^0
\to \gamma \gamma$. Such a result was also found in Ref.~\cite{BH88}
using a Euclidean $O(4)$ cut-off. Thus, in the present framework we
have to choose between preserving the anomaly and not obtaining
factorization or vice-versa, i.e. violating the anomaly and
reproducing factorization.

There is a subtlety in the previous reasoning, because the
prescription of not regularizing the abnormal parity contribution to
the action really means that the result is conditionally convergent
and can be given unambiguously if one insists on maintaining vector
gauge invariance~\cite{SR95}. A practical way to implement this fact
is to introduce a gauge invariant regulator and remove it at the end
of the calculation. The triangle graph is linearly divergent, and thus
a regularization must be introduced. If one insists on preserving
vector gauge invariance the regulator must preserve that symmetry, but
then the axial current is not conserved generating the standard chiral
anomaly. The obvious question arises whether the limit $Q^2 \to \infty
$ must be taken before or after removing the cut-off by sending it to
infinity. If one takes the sequence $Q^2 > \Lambda^2 \to \infty $, the
results of Ref.~\cite{DVY02} produce a constant PDA, in agreement with
our low energy calculation.

\subsection{PDA calculation in light-cone coordinates} 

The pion distribution amplitude is defined through 
\begin{eqnarray}
f_\pi p^\mu \varphi_{\pi } (x) = \int \frac{dk^+ dk^- d^2 k_\perp }{2
(2\pi)^4 } \delta \left( k^+ - x p^+ \right) {\rm Tr} \left[ \chi_P
(k) \gamma^\mu \gamma_5 \right]
\end{eqnarray} 
where the PV regularized form the pion BS wave function,
Eq.~(\ref{eq:pion_wf}), is understood.  Formally, in momentum space,
Eq.~(\ref{eq:pda_def}) corresponds to integration over the quark
momenta in the loop integral used in the evaluation of $f_\pi$
Eq.~(\ref{eq:fpi}), but with $k^+ = p^+ x = m_\pi x $ and $k_\perp$
fixed. Thus, with the PV method and after working out the Dirac
traces, we have to compute
\begin{eqnarray}
 \Psi_{\pi } (x, \vec k_\perp ) &=& -\frac{ 2 i N_c M g_{\pi qq} }{
f_\pi} \int \frac{dk^+ dk^- }{(2\pi)^4 } 
\frac{\delta \left( k^+ - x p^+ \right)}{m_\pi^2 x(1-x) }
 \sum_j c_j
\times \\&\times & \frac1{k^- - m_\pi - \frac{ \vec k_\perp^2 + M^2 +
\Lambda_j^2 + i 0^+ }{m_\pi (1-x) } } \, \frac1{k^- - \frac{ \vec
k_\perp^2 + M^2 + \Lambda_j^2 + i 0^+}{m_\pi x } }\nonumber,
\end{eqnarray}
where, again, the location of poles in the $k^-$ variable has been
explicitly displayed. This integral coincides with that found when
computing the PDF (see Eq.~(\ref{eq:LC_int})). Evaluating the $k^-$
integral gives the pion LC wave function in the NJL model with the PV
regularization:
\begin{eqnarray} 
\Psi_{\pi } (x, k_\perp)= \frac{4 N_c M g_{\pi qq} }{16\pi^3 f_\pi}
\sum_j  c_j \frac1{k_\perp^2 + \Lambda_j^2 + M^2-x(1-x)m_\pi^2 } .
\end{eqnarray} 
For $m_\pi \neq 0 $ it is non-factorizable in the $k_\perp$ and $x$
variables.  As a consequence of the PV condition with two subtractions one has,
for large $k_\perp$,
\begin{eqnarray} 
\Psi_{\pi} (x, k_\perp)& \to & \frac{4 N_c M g_{\pi qq}}{16\pi^3 f_\pi
} \frac{ \sum_i c_i \Lambda_i^4 }{k_\perp^6},
\end{eqnarray} 
which gives a finite normalization and a finite second transverse
moment,
\begin{eqnarray} 
\langle k_\perp^2 \rangle &=& \int d^2 k_\perp \int_0^1 dx \,
\Psi_{\pi} (x, k_\perp) k_\perp^2 \, .
\end{eqnarray} 
Integrating with respect to $k_\perp$ reproduces the pion distribution
amplitude of Eq.~(\ref{eq:pda}). In the chiral limit, $m_\pi = 0$, one
can use the Goldberger-Treiman relation for the constituent quarks, $
g_{\pi qq} f_\pi = M$, yielding in
addition to Eq.~(\ref{eq:pd0}) the
very simple formulas
\begin{eqnarray} 
\Psi_{\pi} (x, k_\perp)&=& \frac12 V_\pi ( x, \vec k_\perp ),
\label{eq:lc0} 
\\ \langle \vec k_\perp^2 \rangle_\pi &=& -\frac{M \langle \bar q q
\rangle }{2 f_\pi^2}.
\label{eq:kp0} 
\end{eqnarray} 
where $V_\pi (x, \vec k_\perp)$ is given by Eq.~(\ref{eq:td0}). Note
that these relations are independent of the PV regulators. 

One aspect of regularization should be analyzed here. According to
previous studies~\cite{BL80}, the value of the LC wave function in the
chiral limit is fixed at $\vec k_\perp = 0_\perp$ by the chiral
anomaly, Eq.~(\ref{eq:lc_anom}). In the PV regularized NJL model this
is not the case, since
\begin{eqnarray}
f_\pi \int_0^1 dx \Psi_{\pi} (x, 0_\perp )&=& F(0,0,0) = 
\frac{4 N_c M^2}{16\pi^3 f_\pi^2} \sum_i
c_i \frac1{\Lambda_i^2 + M^2 } 
\end{eqnarray}  
The first two terms in this equation indicate the consistency of our
calculations between the transition form factor and the low energy
matrix element, but confirms the anomaly violation we have referred to
above. 

This is a clear deficiency of the NJL model and its regularization
procedure, and we do not know of any convincing way of avoiding this
problem in this model~\footnote{In the Euclidean version of the model
the accepted regularization prescription is to regularize the real
part of the action (normal parity processes) and not to regularize the
imaginary part of the action (abnormal parity
processes)~\cite{SR95}. Such a prescription agrees with the anomaly
but {\it does not} agree with factorization and produces instead
Eq.~(\ref{eq:phi_unreg}).}. Nevertheless, it can be shown that it is
possible to write down a chiral quark model based on the concept of
spectral regularization~\cite{Ru01} where the proper anomaly is
reproduced and the results $\varphi_\pi (x,\mu_0)=V_\pi (x,\mu_0)/2 =1
$ still hold~\cite{RB02b}.

\begin{figure}[tbp]
\centerline{\epsfig{figure=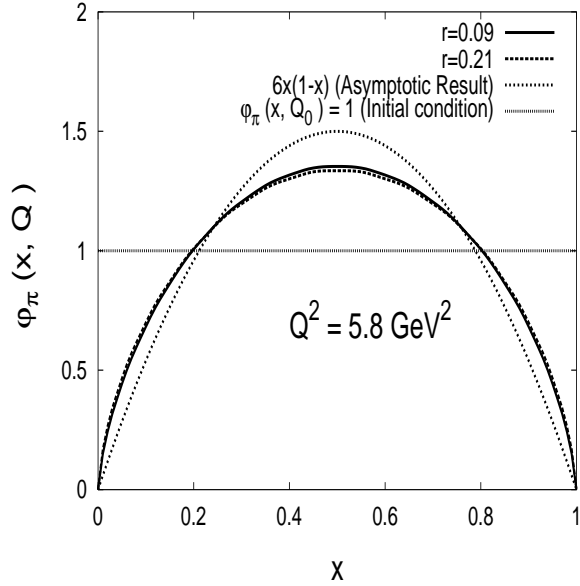,height=8cm,width=8cm}}
\caption{The pion distribution amplitude in the chiral limit evolved
to the scale \protect{$Q^2 = (2.4 {\rm GeV})^2 $}.  The two values for
the evolution ratio $r=\alpha(Q) / \alpha(Q_0)$ reflect the
uncertainties in the values of Ref.~\cite{SY00} based on an analysis
of the CLEO data.  We also show the unevolved PDA,
$\varphi_\pi(x,Q_0)=1$, and the asymptotic PDA, $\varphi_\pi
(x,\infty)=6x(1-x)$. We use, as suggested by the analysis of the DIS,
$\alpha(Q_0)=2.14 $ of Ref.~\cite{DR95}, which corresponds to
$Q_0=313$~MeV. }
\label{fig:pda}
\end{figure}

\subsection{QCD evolution} 

The leading-twist PDA requires, the inclusion of
radiative logarithmic corrections through the QCD
evolution~\cite{BL80}. For the pion distribution amplitude this
is done in terms of the Gegenbauer polynomials, by interpreting our
low-energy model result as the initial condition. 
In the chiral limit, 
\begin{eqnarray}
\varphi_{\pi} (x,\mu_0) = 1.
\label{start}
\end{eqnarray}
Then, the LO-evolved distribution amplitude reads~\cite{BL80}
\begin{eqnarray}
\varphi_{\pi} (x,\mu) &=& 6x(1-x){\sum_{n=0}^\infty}'  C_n^{3/2} ( 2 x -1)
a_n (\mu),
\label{eq:evolpda} 
\end{eqnarray}
where the prime indicates summation over even values of $n$ only. The matrix
elements, $a_n(\mu)$, are the Gegenbauer moments given by
\begin{eqnarray}
a_n (\mu)&=& \frac23 \frac{2n+3}{(n+1)(n+2)} \left(
 \frac{\alpha(\mu_{})}{\alpha(\mu_0) } \right)^{\gamma_n^{\rm NS} / (2
 \beta_0)} \times \nonumber \\ &&\int_0^1 dx C_n^{3/2} ( 2x -1)
 \varphi_{\pi} (x ,\mu_0),
\label{Geg}
\end{eqnarray}
with $C_n^{3/2}$ denoting the Gegenbauer polynomials, and
$\gamma_n^{\rm NS} > 0 $ is given by Eq.~(\ref{eq:anom_dim}).  From
Eq.~(\ref{start}) one gets immediately
\begin{eqnarray}
\int_0^1 dx C_n^{3/2} ( 2x -1) \varphi_{\pi} (x ,Q_0) =1.
\label{ourGeg}
\end{eqnarray} 
Thus, for a given value of $\mu$ we may predict PDA.  To determine the
initial scale $\mu_0$, or, equivalently, the evolution ratio
$r=\alpha(Q) / \alpha(Q_0 )$ Ref.~\cite{RB02} uses the result of
Ref.~\cite{SY00}, where it is found $a_2 (2.4 {\rm GeV}) = 0.12 \pm
0.03$. Using this input, one gets
\begin{eqnarray} 
\mu_0 = 313^{+60}_{-30} {\rm MeV}
\label{eq:m0_pda} 
\end{eqnarray} 
Within uncertainties, this result is compatible with the values
obtained from the momentum fraction analysis, Eq.~(\ref{eq:mu0_dis})
and the pion electromagnetic form factor, Eq.~(\ref{eq:mu0_piff}).  The
result for the pion distribution amplitude obtained in
Ref.~\cite{RB02} is shown in Fig.~\ref{fig:pda} for $Q = 2.4 {\rm GeV}
$ and reflecting the uncertainties from Ref.~\cite{SY00}. After
evolution the results closely resemble those found in the transverse
lattice~\cite{Da01,BS01,BD02}. The analysis of the end point behavior 
yields, {\it after } evolution, the estimate  
\begin{eqnarray} 
\varphi_\pi (x, \mu ) \to 8 x \zeta \left( \frac{4 C_F }{\beta_0} \ln
\frac{\alpha(\mu)}{ \alpha(\mu_0)} +1 \right) \sim \, 12.5 \, x \, , 
\end{eqnarray} 
for $ x\to 0 $. Here $\zeta(s)= \sum_{n=1}^\infty n^{-s}$ is the
Riemann zeta function and the numerical value corresponds to take $\mu
= 2.4 {\rm GeV} $. The value obtained in Ref.~\cite{RB02} for the
second $\xi$-moment ($\xi = 2x-1$) is
\begin{eqnarray}
\langle \xi^2 \rangle &=& \int_0^1 dx \, \varphi_\pi (x, Q=2.4 {\rm
GeV} ) (2x-1)^2 
 \nonumber \\ &=& 0.040 \pm 0.005,
\end{eqnarray} 
to be compared with $ \langle \xi^2 \rangle = 0.06 \pm 0.02 $ obtained
in the standard lattice QCD for $Q= 1/a = 2.6 \pm 0.1 {\rm
GeV}$~\cite{DP00}. 

Based on the identities $\varphi_\pi (x,\mu_0) = V_\pi (x,\mu_0)/2= 1
$ between the PDF and the PDA in the chiral limit at the quark model
scale a remarkable integral relation between $\varphi_\pi (x,\mu) $
and $ V_\pi (x,\mu)/2$ has recently been derived using LO evolution
equations~\cite{RB02},
\begin{eqnarray}
\frac{\varphi_\pi (x, \mu) }{6 x (1-x) }-1 = \int_0^1 dy K(x,y) V_\pi
(y,\mu)  
\label{eq:int_eq} 
\end{eqnarray}  
where the explicit expression for the scale independent kernel,
$K(x,y)$ is given in Ref.~\cite{RB02}. Using this equation one can
regard Fig.\ref{fig:pda} as a prediction of the PDA in terms of the
PDF parameterizations of Ref.~\cite{SMRS92} (see
Fig.\ref{fig:pionval}). Using Eq.~(\ref{eq:int_eq}) one gets the
following estimate for the leading twist contribution to the pion form
factor at LO 
\begin{eqnarray}
\frac{Q^2 F_{\gamma^* , \pi \gamma} (Q) }{2 f_\pi} \Big|_{\rm Twist-2} = 
\int_0^1 dx \frac{\varphi_\pi (x,Q )}{6x(1-x)} = 1.25 \pm 0.10  
\end{eqnarray}  
The experimental value obtained in CLEO~\cite{CLEO98} for the full
form factor is $ Q^2 F_{\gamma^* , \pi \gamma} (Q) / 2 f_\pi = 0.83
\pm 0.12 $ at $Q^2 = 5.8 {\rm GeV}^2 $. Taking into account that we
have not included neither NLO effects nor an estimate of higher twist
contributions, the result is rather encouraging. Finally, using this
estimate for $\varphi_\pi (x,Q)$ in the electromagnetic form factor,
Eq.~(\ref{eq:pion_ffpda}), one deduces an enhacement 
\begin{eqnarray}
-Q^2 F_\pi^{\rm em} (-Q^2) \Big|_{\rm Twist-2}   
= 0.20 \pm 0.03    \quad \, \qquad Q^2
= 5. 8 {\rm GeV}^2 
\end{eqnarray} 
which also brings the number closer to experimental number~\cite{FF},
although it only accounts for half of its value $ \sim 0.38 \pm 0.04
{\rm GeV}^2 $, perhaps due to the lack of NLO or higher twists. At
present this point seems not to be understood and deserves further
investigation.

\section{Conclusions}

Chiral quark models incorporate two essential features of QCD at low
energies: spontaneous chiral symmetry breaking and quark degrees of
freedom. In addition, it can be made compatible with large $N_c$
counting rules, and calculations in the leading order approximation
have been undertaken in the past. In these lectures we have focused
and reviewed particular applications in a prototype chiral quark
model, the Nambu--Jona-Lasinio model, but many results extend
trivially to other models. We have also restricted to the pion because
we do not expect a better theoretical understanding of a hadron,
including the fact that confinement is hoped not to play an essential
role. 

A very important issue concerning the treatment of low energy models
is the existence of a high momentum suppression in the interaction. In
the NJL model this is done via a regularization method, which has to
comply with several properties, like gauge invariance and scaling in
the high energy limit, i.e. the absence of spurious logarithmic
corrections. This allows to clearly identify power corrections of pion
observables.

These models make sense at a given low renormalization scale
$\mu_0$. Any result for a given observable may be used to compute that
observable at a higher scale $\mu$ through QCD evolution. In this way
the correct behavior for QCD radiative corrections may be incorporated
using the chiral quark model as an initial condition for the
evolution. Although radiative corrections are only known
perturbatively, there is a chance that the matching scale, $\mu_0$,
makes perturbation theory meaningful. This a weak point which can only
be addressed by computing higher order corrections to the model and
higher order corrections to the QCD evolution. Nevertheless, it is
encouraging that three different determinations of the scale based on
matching to LO perturbative QCD evolution and available experimental
data yield within uncertainties the scale $\mu_0 = 320 {\rm MeV} $ for
$ \Lambda_{\rm QCD} = 225 {\rm MeV} $. Actually, using this low scale
the description of the distribution functions and distribution
amplitudes agrees remarkably well with phenomenological analysis.

\section*{Acknowledgments}
Much of these lectures is based on common work and discussions with
R. D. Davidson, H. Weigel, L. Gamberg and W. Broniowski. I also thank
M. Praszalowicz for discussions. Finally, I wish to thank the
organizers for the kind invitation and the pleasent atmosphere.
Support from DGES (Spain) Project PB98-1367 and by the Junta de
Andaluc\'\i a is acknowledged.  Partial support from the Spanish
Ministerio de Asuntos Exteriores and the Polish State Commitee for
Scientific Research, grant number 07/2001-2002 is also gratefully
acknowledged.

\def\NCA{{\em Nuovo Cimento}  A }
\def\NIM{\em Nucl. Instrum. Methods }
\def\NIMA{{\em Nucl. Instrum. Methods} A }
\def\NPB{{\em Nucl. Phys.} B }
\def\NPA{{\em Nucl. Phys.} A }
\def\PLB{{\em Phys. Lett.}  B }
\def\PRL{{\em Phys. Rev. Lett.} }
\def\PRD{{\em Phys. Rev.} D }
\def\PRC{{\em Phys. Rev.} C }
\def\ZPC{{\em Z. Phys.} C }
\def\JPA{{\em J. Phys.} A }
\def\EPJC{{\em Eur. Phys. J.} C }

\newpage 


\begin{thebibliography}{99}
\bibitem{Yn83} See, for instance F. J. Yndurain, {\it Quantum
Chromodynamics}, Springer Verlag, Berlin 1893.

\bibitem{Pe79} {For a review see e.g., A. Peterman, Phys. Rep. {\bf
53} (1979) 157. } 

\bibitem{AP77}
G. Altarelli and G. Parisi, \NPB {\bf   126} (1977) 298. \\
For reviews see, e.g. \\ 
A. Buras, Rev. Mod. Phys. {\bf 52} (1980) 199, \\ 
E. Reya, Phys. Rep. {\bf 69} (1981) 195. 



\bibitem{BL80} 
G. P. Lepage and S. J. Brodsky, Phys. Lett. {\bf B 87}  (1979) 359. \\
G. P. Lepage and S. J. Brodsky, Phys. Rev. {\bf D 22} (1980) 2157. \\ 
D. M\"uller, Phys. Rev. {\bf D 51} (1995) 3855. 

\bibitem{GL84} 
J. Gasser and H. Leutwyler, Ann. Phys.{\bf 158} (1984) 142. \\ 
J. Gasser and H. Leutwyler, \NPB {\bf .B250} (1985)465.

\bibitem{Pi95} For a review see e.g. A. Pich, Rep. Prog. Phys. {\bf
58} (1995) 563. 

\bibitem{AS01} D. Arndt and M. Savage, \NPA {\bf 697} (2002) 429.

\bibitem{NJL61} {Y. Nambu and G. Jona-Lasinio, Phys. Rev. {\bf 122} (1961)
                    345; {\bf 124} (1961) 246.}

\bibitem{NJL} {For reviews see, e.g., \\ U. Vogl and W. Weise, {\it
Prog. Part. Nucl. Phys.} vol. {\bf 27} (1991) 195; \\ S. P. Klevansky,
Rev. Mod. Phys. {\bf 64} (1992) 649; \\ M. K. Volkov, Part. and Nuclei
{\bf B 24} (1993) 1; \\ T. Hatsuda and T. Kunihiro, Phys. Rep. {\bf
247} (1994) 221; \\ Chr. V. Christov, A. Blotz, H.-C. Kim,
P. Pobylitsa, T. Watabe, T. Meissner, E. Ruiz Arriola, and K. Goeke,
Prog. Part. Nucl. Phys. {\bf 37} (1996) 91; \\ R. Alkofer,
H. Reinhardt, and H. Weigel, Phys. Rept. {\bf 265} (1996) 139; \\
G. Ripka, {\it Quarks Bound by Chiral Fields} Oxford Science
Publications, 1997, and references therein. }


\bibitem{JR80} R. L. Jaffe and G. C. Ross, \PLB {\bf 93} (1980) 313.

\bibitem{Ja85} For a review see e.g. R. L. Jaffe, {\it Relativistic
Dynamics and Quark Nuclear Physics}, proceedings of the Los Alamos
School, 1985, edited by M. B. Johnson and A. Picklesimer ( Wiley, New
York, 1986).

\bibitem{PDG} Review of Particle Physics, K. Hagiwara et al.,
Phys. Rev. {\bf D 66} (2002) 010001. 

\bibitem{SMRS92} P.J. Sutton, A.D. Martin, R. G. Roberts and W.J.
Stirling, \PRD {\bf 45} (1992) 2349.

\bibitem{Ru98a} E. Ruiz Arriola, \NPA {\bf  641} (1998) 461. 

\bibitem{Co01} A. Cooper-Sarkar, Contribution to International
Europhysics Conference on High-Energy Physics (HEP 2001), Budapest,
Hungary, 12-18 Jul 2001.  Published in {\it Budapest 2001, High energy
physics} hep2001/009. 

\bibitem{BR93} R. D. Ball and G. Ripka in Proceedings of {\it Conference on
Many-Body Physics} Coimbra, Portugal, 20-25 Sep
1993. Eds. C. Fiolhais, M. Fiolhais, C. Sousa and J. N. Urbano 
(World Scientific, 1994). hep-ph/9312260

\bibitem{Do92} F. D\"oring, A. Blotz, C. Sch\"uren, T. Meissner,
E. Ruiz Arriola and K. Goeke; \NPA {\bf 536}(1992) 548.


\bibitem{DS95} V. Dmitrasinovic, H.J. Schulze, R. Tegen and
R. H. Lemmer; Ann. Phys. {\bf 238} (1995) 332. 

\bibitem{NB96} E. N. Nikolov, W. Broniowski, C. V. Christov, G. 
Ripka, K. Goeke \NPA {\bf 608} (1996) 411. 

\bibitem{OB01} For a review see e.g. M. Oertel, M. Buballa and J. Wambach,
Phys.Atom. Nucl. {\bf 24} (2001) 757 and references therein. 

\bibitem{PV49} 
W. Pauli and F. Villars, Rev. Mod. Phys. {\bf 21}
                 (1949) 434. 

\bibitem{Ru91} E. Ruiz Arriola, Phys. Lett. {\bf B 253} (1991) 430. 

\bibitem{CR91} C.V. Christov, E. Ruiz Arriola and K. Goeke, Acta
Phys. Polon. {\bf B 22} (1991) 187.

\bibitem{SR92} {C. Sch\"uren, E. Ruiz Arriola, and K. Goeke,
Nucl. Phys. {\bf A 547} (1992) 612.}

\bibitem{DR95} R. M. Davidson and E. Ruiz Arriola, Phys. lett. {\bf B
348} (1995) 163.


\bibitem{DR96} R. M. Davidson and E. Ruiz Arriola, \PLB {\bf 359}
(1995) 273. 


\bibitem{WRG99} H. Weigel, E. Ruiz Arriola, and L. P. Gamberg,
Nucl. Phys. {\bf B 560} (1999) 383.

\bibitem{DR02} R. M. Davidson and E. Ruiz Arriola,
Act. Phys. Pol. {\bf B 33} (2002) 1791. 

\bibitem{RB02} 
E. Ruiz Arriola and W. Broniowski, \PRD (in press). hep-ph/0207266 

\bibitem{CK01} For a recent review see, {\em e.g.}, P. Colangelo and
A. Khodjamirian, in {\it Handbook of QCD}, ed. by M. Shifman (World
Scientific, Singapore, 2001) vol. {\bf 3} p. 1495. 

\bibitem{Io02} B.L. Ioffe;  hep-ph/0207191.  

\bibitem{IZ02} B.L. Ioffe, and K.N. Zyablyuk.  

\bibitem{FF} C. J. Bebek {\it et al.}, \PRL {\bf 37} (1976). \\ 
For review see e.g. H. P. Blok, G. M. Huber and D. J. Mack,
nucl-ex/0208011.

\bibitem{FJ79} G. F. Farrar and D. R. Jackson, \PRL {\bf 43} (1979)
246. 

\bibitem{BH88} A. Blin, B. Hiller, and M. Schaden, Zeit. Phys. {\bf A
331} (1988) 75. 

\bibitem{Co89} J. S. Conway {\it et al.} \PRD {\bf 39} (1989) 92.

\bibitem{Au89} P. Aurenchie, R. Baier, M. Fontanaz,
M.N. Kienzle-Focacci and M. Werlen \PLB {\bf 233} (1989) 517. 

\bibitem{kla01} M. Klasen,  J. Phys. {\bf G 28} (2002) 1091. 

\bibitem{bf73} S. J. Brodsky and G. R. Farrar, Phys. Rev. Lett. {\bf
31} (1973) 1153; Phys. Rev. {\bf D 11}(1975) 1309.

\bibitem{Me02} W. Melnitchouk, hep-ph/0208258.

\bibitem{JM97} R. Jakob, P.J. Mulders, J. Rodrigues, 
\NPA {\bf  626} (1997) 937.

\bibitem{BH99}  W. Bentz, T. Hama, T. Matsuki and K. Yazaki, 
\NPA {\bf 651} (1999) 143. 

\bibitem{SS93} T. Shigetani, K. Suzuki and  H. Toki, \PLB {\bf 308}
(1993) 383; \NPA {\bf A 579} (1994) 413. 


\bibitem{FM94} T. Frederico and G. A. Miller, \PRD {\bf  50}
(1994) 210. 


\bibitem{LB81} G.P. Lepage, S. J. Brodsky, T. Huang and P.B.
Mackenzie.  Invited talk given at Banff Summer Inst. on Particle
Physics, Banff, Alberta, Canada, Aug 16-28, 1981.

\bibitem{St99} G. Stoll, hep-ph/9812432 


\bibitem{Br01} For a review see e.g. S. J. Brodsky, Invited talk at
Kracow School of Theoretical Physics: 41st Course: Fundamental
Interactions, Zakopane, Poland, 2-11 Jun 2001.  Acta
Phys. Polon. {\bf B32} (2001) 4013.

\bibitem{RB02b} E. Ruiz Arriola and W. Broniowski, (work in
preparation).





\bibitem{SW95} C.M. Shakin, Wei-Dong Sun, \PRC {\bf 51} (1995) 2171.


\bibitem{BC02} F. Bissey, J.R. Cudell, J. Cugnon, M. Jaminon,
J.P. Lansberg and P. Stassart, hep-ph/0207107

\bibitem{DT00} A.E. Dorokhov and L. Tomio, \PRD {\bf  62} (2000)
014016. 

\bibitem{HRS01} M.B. Hecht, C.D. Roberts and  S.M. Schmidt,  
\PRC {\bf  63} (2001) 025213.  

\bibitem{Ru01} E. Ruiz Arriola, Proceedings of the Workshop on {\it
Lepton Scattering, Hadrons and QCD}, Adelaide, (Australia)
2001. Published by World Scientific, eds. W. Melnitchouk,
A. Schreiber, P. Tandy and A. W. Thomas.






\bibitem{GRS98} M. Gluck, E. Reya and M. Stratmann, \EPJC {\bf 2}
(1998) 159.

\bibitem{GRS99} M. Gluck, E. Reya, I. Schienbein, \EPJC {\bf
10} (1999) 313.


\bibitem{AP96} G. Altarelli, S. Petrarca and F. Rapuano, \PLB {\bf B
373} (1996) 200.

\bibitem{Ru98b} E. Ruiz Arriola, Talk given at Miniworkshop on Hadrons
as Solitons, Bled, Slovenia, 1999, {\it hep-ph/9910382}.










\bibitem{CLEO98} CLEO Collaboration (J. Gronberg et al.),
Phys. Rev. {\bf D57} (1998) 33. 


\bibitem{SY00} A. Schmedding and O. Yakovlev, Phys. Rev. {\bf D 62} (2000)
116002.


\bibitem{CZ84} V. L. Chernyak and  A. R. Zhitnitsky, Phys. Rept. {\bf
112} (1984) 173. 


\bibitem{BJ97} V. M. Belyaev and M. B. Johnson, Phys. Rev. {\bf D
56} (1997) 1481. 

\bibitem{DP00} L. Del Debbio, M. Di Pierro, A. Dougal and
C. Sachrajda, Nucl. Phys. {\bf B} (Proc. Suppl.) {\bf 83-84 } (2000) 235. 

\bibitem{Da01} S. Dalley, Phys. Rev. {\bf D 64} (2001) 036006.

\bibitem{BS01} M. Burkardt and S.K. Seal, Phys. Rev. {\bf D 65} (2002)
034501.

\bibitem{BD02} M. Burkardt and S. Dalley, hep-ph/0112007.  

\bibitem{PP97} V. Yu. Petrov and P. V. Pobylitsa, hep-ph/9712203. 

\bibitem{PP99} V. Yu. Petrov, M. V. Polyakov, R. Ruskov, C. Weiss, and
K. Goeke, Phys. Rev. {\bf D 59} (1999) 114018. 

\bibitem{ADT00} I. V. Anikin, A. E. Dorokhov and L. Tomio,
Phys. Lett. {\bf B 475} (2000) 361. 

\bibitem{He00} T. Heinzl, Nucl. Phys. {\bf B} (Proc. Suppl.) 90 (2000)
83.

\bibitem{He01} T. Heinzl, Lect. Notes Phys. {\bf 572} (2001) 55.
hep-th/0008096.

\bibitem{PR01} M. Prasza\l{}owicz and A. Rostworowski, Phys. Rev. {\bf D
64} (2001) 074003.

\bibitem{ADT01} I. V. Anikin, A. E. Dorokhov, and L. Tomio,
Phys. Atom. Nucl. {\bf 64} (2001) 1329.

\bibitem{DVY02} A. E. Dorokhov, M. K. Volkov, and V. L. Yudichev,
hep-ph/0203136.

\bibitem{Do02} A. E. Dorokhov, talk presented at the 37th Rencontres
de Moriond on QCD and Hadronic Interactions, Les Arcs, France, 16-23
March 2002, hep-ph/0206088.


\bibitem{SR95} E. Ruiz Arriola and L. L. Salcedo, Nucl. Phys. {\bf A
590} (1995) 703. 





\end{thebibliography}
\end{document}